\documentstyle[aps,epsf,aps10,axodraw_3,graphicx,floats]{revtex}
\begin{document}

\def\vq1{\slash{q}_1}
\def\bok{\slash{k}}
\def\beq{\begin{equation}}
\def\eeq{\end{equation}}
\def\beqy{\begin{eqnarray}}
\def\eeqy{\end{eqnarray}}
\newcommand{\qk}{quark}
\newcommand{\gls}{gluons}

\newcommand{\bleft}[1]{\left#1 \rule{0em}{1.25em} \right.}
\newcommand{\bright}[1]{\left. \rule{0em}{1.25em} \right#1}

\title{A Phenomenological Lagrangian Approach to Two Kaon Photoproduction and 
Pentaquark Searches}
\author{W. Roberts}
\address{Department of Physics, Old Dominion University, Norfolk, VA
23529, USA\\
and\\
 Thomas Jefferson National Accelerator Facility,
12000 Jefferson Avenue, Newport News, VA 23606, USA.}
\date{\today}
\maketitle
\begin{abstract}
We examine cross sections for the processes $\gamma N\to NK\overline{K}$ in the
framework of a phenomenological Lagrangian. We include contributions from
$\Lambda$ and $\Sigma$ resonances up to spin 3/2, as well as those from an
exotic $\Theta^+$. We allow the $\Theta^+$ to have spin 1/2 or 3/2, with either
positive or negative parity in each case. We also allow the state to be either
isovector or isoscalar. We find that the scenario that most closely matches
observations at Jefferson Laboratory requires a moderately large coupling of the
$\Theta^+$ to $NK^*$.  
\vspace{5mm}
 
\flushright{JLAB-THY-04-256}
\end{abstract}
\thispagestyle{empty}
\pacs{\tt$\backslash$\string pacs\{13.60.-r, 13.60.Rj,13.60.Le\}}
\setcounter{page}{1}
\def\slash#1{#1 \hskip -0.5em / }

\section{Introduction}

\subsection{Experimental Indications}

In the past several months, a number of experimental groups have reported
signals for a pentaquark state called the $\Theta^+$ \cite{spring8} --
\cite{cosy}.  The first evidence for such a state was reported by the Spring-8
Collaboration \cite{spring8}. The search by Spring-8 was motivated by
predictions, made within the framework of the chiral soliton model, by Diakonov
{\it et al.} \cite{diakonov}. Most searches that have reported evidence for the
state put its mass around 1540 MeV. However, in all cases, the experimental
resolution has been such that only upper limits for the width  of the state
could be given. Evidence for other pentaquarks predicted as partners to the
$\Theta^+$, particularly the $\Xi^{--}$, has also been reported by the NA49
Collaboration \cite{na49}. Using the time-delay technique, Kelkar {\it et al.}
have found evidence of not only the $\Theta^+$ in $K^+N$ scattering data, but
also a possible spin-orbit partner \cite{kelkar} along with a third possible
state.

Despite the number of pentaquark sightings, the situation is far from clear.
Two members of the NA49 Collaboration produced a minority report pointing out
that there was no strong evidence for the existence of the $\Xi^{--}$ in older,
higher-precision data \cite{fischer}. The HERA-B Collaboration sees no evidence
for the $\Theta^+$ \cite{herab}, and the BES Collaboration also report no
evidence in their searches \cite{bes}. Searches at RHIC have also yielded no
evidence so far \cite{rhic}. Because some experiments have reported signals for
the pentaquarks, while others have seen no sign of them, Karliner and Lipkin
\cite{lipkinc} have postulated the existence of a `crypoexotic' $N^*$ that
plays a significant role in production of the $\Theta^+$. In addition, none of
the experiments that report a signal for any of the pentaquarks can say
anything about their spin or parity.

Quite apart from the question of the existence of the $\Theta^+$, the question
of its width is also very interesting. Nussinov \cite{nussinova} has examined
the implication of such a state for existing $K^+d$ data, and has concluded
that the width of the state had to be less than 6 MeV. Arndt and collaborators
\cite{workmana} have performed a similar analysis on $K^+N$ scattering data,
and have concluded that the width has to be less than 1 MeV, while Haidenbauer
and Krein \cite{haidenbauer} conclude that the width of the state must be less
than 5 MeV, or that its mass must be much lower than reported. Cahn and
Trilling \cite{cahn} have suggested that the width is $0.9\pm 0.3$ MeV, based
on their analysis of data from $K^+$ collisions on xenon. Gibbs \cite{gibbs}
has also examined $K^+d$ data and has extracted a width of $0.9\pm 0.2$ MeV.

\subsection{Theoretical Implications}

The existence of a pentaquark state wouldn't be too jarring for most QCD
practitioners, as multiquark states have been anticipated for decades. However,
its light mass and apparently narrow width are difficult to explain in a
`conventional' scenario, and have stimulated much discussion and many
postulates. Dzierba {\it et al.} have raised the possibility that the `signal'
is really a kinematic reflection \cite{dzierba}. Jaffe and Wilczek \cite{JW}
have constructed a diquark scenario for the $\Theta^+$. One consequence of
their scenario is that the state should have a spin-orbit partner, for which
there is little or no evidence to date. Capstick and collaborators \cite{cpr}
have suggested that the state is as narrow as it is because it has isospin 2.
This means that there should be isospin partners, none of which have been seen.

Jennings and Maltman \cite{maltman} have examined pentaquark phenomenology in
a number of scenarios, and conclude that such a state fits into the quark
model picture if its parity is positive, but this implies the existence of
spin-orbit partners. Karliner and Lipkin \cite{lipkin} invoke the mixing of two
nearly-degenerate $KN$ resonances to explain the narrow width of the
$\Theta^+$. They have also speculated on the phenomenology of pentaquark states
containing a charm quark \cite{lipkinb}. In addition, there are many papers
that examine the phenomenology of  pentaquarks using QCD sum rules \cite{sum},
various quark models \cite{oha} and string theory \cite{bando}. A number of
unique scenarios have also been proposed \cite{unique}. There have  even been
suggestions that the states seen are in fact heptaquarks \cite{bicudo}, or
$NK\pi$ bound states \cite{nkp}. Some lattice simulations suggest that the
parity of the state is negative \cite{lattice,sasaki}, while the work of Chiu
and Hsieh suggest that it is positive \cite{latticeb}. More recent lattice work
reports no signal for the pentaquark state \cite{latticec}.

\subsection{Cross Section Ramifications}

Among the many unanswered questions regarding the $\Theta^+$ (and other
pentaquark candidates) is that of the cross section for its production in a particular
reaction. To address this, a number of authors have examined cross sections for
producing them in a number of reactions \cite{pr}-\cite{n-t}. Most of these
have been aimed at determining the spin and parity of the state, but they have
all provided estimates for the production cross section. Various
photoproduction mechanisms and observables are examined in refs. \cite{pr} -
\cite{n-t}, while the authors of ref. \cite{hyodo} treat the reaction
$K^+p\to\pi^+K^+n$ with kinematics suited to production of the $\Theta^+$. The
authors of ref. \cite{oh} also examine production of the $\Theta^+$ using a
pion as the incident particle.

The calculation of ref. \cite{n-t} is closest in spirit to the work that we
present here. Those authors use a phenomenological Lagrangian to describe the
reaction $\gamma n \to nK^+K^-$. In addition to the contribution of the
$\Theta^+$, they also included two $\Sigma^-$ hyperons: the $\Sigma^-(1197)$
and the $\Sigma^-(1660)$. These hyperons provided the background contribution in
their calculation. Close and Zhao \cite{close} have emphasized the importance
of comparing the cross section for producing the pentaquark states with those
for producing non-exotic hyperons.

In this manuscript, we examine the cross section for the process $\gamma N\to
NK\overline{K}$. We include many contributions, and examine all of
the channels that are allowed. By including a number of contributions, we are
able to understand the roles played by non-exotic hyperons, and by the
$\phi(1020)$. More specifically, since some of the properties of the non-exotic
hyperons are known, these can be used to get a handle on how big are the cross
sections for their production, and for the production of the $\Theta^+$.

The results of this calculation are relevant to past, presnt and future
searches using photon beams. For the published results so far, this means the
searches at JLab \cite{clas,clas-2}, the search by the Saphir Collaboration
\cite{saphir} and, of course, the search by the Spring-8 Collaboration
\cite{spring8}. We note, however, that in ref. \cite{clas-2}, the process
studied is $\gamma p\to n\pi^+K^+K^-$, not any of the ones discused in this
manuscript. While the main focus of our discussion will be the JLab searches,
we will also comment on the other two searches where appropriate.

The rest of this article is organized as follows. The next two sections focus
on establishing the framework for the calculation: the general amplitude,
kinematics and cross section are discussed in the next section, and the
phenomenological Lagrangian terms and most of the coupling constants needed for
building the model are presented in section III. The diagrams representing the
contributions that are included in this calculation are also shown in that
section. We present our results in section IV, and a summary and outlook in
section V.

\section{General Amplitude, Kinematics and Cross Section}

\subsection{Kinematics and Cross Section}

We begin by describing the kinematics of the process. $k$  is the momentum of the photon,
$p_1$ is that of the target nucleon, $p_2$ is that of the scattered nucleon, 
and  $q_1$ and $q_2$ are the kaon momenta. Momentum conservation gives
\begin{equation}
k+p_1=p_2+q_1+q_2.
\end{equation}
This means that when we construct the amplitude for the process using all the 
four-vectors at our disposal, we can eliminate one of these from consideration. 

The total center-of-mass (com) energy of the process is $\sqrt{s}$, where
$s=(k+p_1)^2$. We may define a variable $t$ as the square of the momentum
transfered to the nucleon, namely, $t=(p_2-p_1)^2$, and this is related to the
scattering angle of the nucleon in the c-o-m frame.

The differential cross section for this process is described in terms of five
kinematic variables. These may be, for instance, two Lorentz invariants and
three angles. One obvious choice for one of the invariants is $s$. The choice
of the other four quantities can be fairly arbitrary, and depends on what
information is being presented. One choice is the scattering angle of the 
nucleon, $\theta$, or equivalently, $t$. For the other three variables, we can
choose for example, $s_{K\overline{K}}\equiv (q_1+q_2)^2$ and
$d\Omega_{K\overline{K}}^*\equiv d\Theta_{K\overline{K}}^*d
\Phi_{K\overline{K}}^*$. Here, $\Theta^*_{K\overline{K}}$ and 
$\Phi_{K\overline{K}}^*$ are determined in the rest frame of the
$K\overline{K}$ pair, relative to a $z^\prime$ axis defined by the direction
of motion of the pair of kaons. Another equally valid choice would be
$s_{NK}\equiv(p_2+q_1)^2$ and $d\Omega_{NK}^*$, where the solid angle is
defined in the rest frame of the nucleon-kaon pair.

The differential cross section is
\begin{equation}
\frac{\partial\sigma}{\partial M_{NK}^2\partial M_{N\overline{K}}^2}
=\frac{1}{(2\pi)^5}\frac{1}{p_1\cdot k}\int\overline{\left|M\right|^2} 
d\cos{\theta} d\Phi_{K\overline{K}}^*,
\end{equation}
where $\theta$ is the scattering angle of the $K\overline{K}$ pair relative to
the momentum of the incident photon, in the rest frame of the initial photon and
nucleon.

\subsection{General Amplitude}

Our starting point is the construction of the most general form for the
transition amplitude for this process. While the requirements of Lorentz
covariance and gauge invariance delimit the form of the amplitude, we find that
there is nevertheless quite a bit of freedom in the form chosen. The most 
general form is 
\begin{equation}
i {\cal M}=\overline{U(p_2)}\varepsilon_{\mu} {\cal O}^\mu U(p_1)
\end{equation}
where
\begin{eqnarray} \label{eq:Omu}
{\cal O}^\mu &=& a_1 p_1^\mu + a_2 p_2^\mu + a_3 q_1^\mu + a_4 \gamma^\mu
+\slash{k}\left(a_5 p_1^\mu + a_6 p_2^\mu + a_7 q_1^\mu + a_8 \gamma^\mu\right)
\nonumber \\ 
&+& \slash{q}_1 \left(a_9 p_1^\mu + a_{10} p_2^\mu +a_{11}q_1^\mu +a_{12} 
\gamma^\mu\right)
+\slash{q}_1 \slash{k}\left(a_{13} p_1^\mu + a_{14} p_2^\mu +a_{15}q_1^\mu+
a_{16} 
\gamma^\mu \right).
\end{eqnarray}
Note that we have no terms in $\slash{p}_1$ nor $\slash{p}_2$, as the initial 
and final nucleons each satisfy
\begin{equation}
\slash{p}U(p)=m U(p).
\end{equation}

The amplitude coefficients $a_i$ are all functions of the kinematic variables
$s$, $s_{K\overline{K}}$, $\theta$, $\Theta^*$ and $\Phi^*$, or whatever
combination of kinematic variables is chosen. Their exact dependence on each of
these variables will be determined by the specific model constructed.

Gauge invariance of the amplitude requires that $k_\mu {\cal O}^\mu=0$, which 
leads to the four relations
\begin{eqnarray}
&&a_1k\cdot p_1 +a_2 k\cdot p_2 +a_3 q_1\cdot k=0,\\
&&a_4+a_5k\cdot p_1 +a_6 k\cdot p_2 +a_7 q_1\cdot k=0,\\
&&a_9k\cdot p_1 +a_{10} k\cdot p_2 +a_{11} q_1\cdot k=0,\\
&&a_{12}+a_{13}k\cdot p_1 +a_{14} k\cdot p_2 +a_{15} q_1\cdot k=0.
\end{eqnarray}
Note that there is no condition on either $a_8$ or 
$a_{16}$.

From these equations, we can eliminate four of the amplitude coefficients,
leaving us with twelve independent ones, or Lorentz-Dirac structures,
to describe the amplitude. One choice would be to eliminate $a_1$, $a_4$,
$a_9$, $a_{12}$, which gives
\begin{eqnarray} 
\varepsilon_\mu {\cal O^\mu}&=&\left\{
\frac{1}{p_1\cdot k}\left[(a_2 + a_{10} \vq1)p_{2\mu} p_{1\nu} +  (a_3 + a_{11}
\vq1)  q_{1\mu} p_{1\nu}\vphantom{\frac{1}{2}}\right]+(a_5 + a_{13}
\vq1)p_{1\nu}  \gamma_\mu \right. \nonumber \\ &+& \left.(a_6 + a_{14} \vq1)
p_{2\mu} \gamma_\nu +(a_7 + a_{15} \vq1)  q_{1\mu}  \gamma_\nu
\vphantom{\frac{1}{p_1\cdot k}}-\frac{1}{2}                    (a_8 + a_{16}
\vq1) \gamma_\mu \gamma_\nu\right\}F^{\mu\nu}, \nonumber   \end{eqnarray}
where $F^{\mu\nu}=\varepsilon^\mu k^\nu-\varepsilon^\nu k^\mu$. Another choice 
is $a_1$, $a_5$, $a_9$, $a_{13}$, giving 
\begin{eqnarray} \varepsilon_\mu {\cal O^\mu} &=&                                                         \left\{
\frac{1}{p_1\cdot k}\left[\vphantom{\frac{1}{p_1\cdot k}}\left(a_2 + \bok a_6+
\vq1 a_{10} +\vq1\bok a_{14}\right)p_{2\mu} p_{1\nu} +\left(a_4 + \vq1
a_{12}\right)p_{1\nu}\gamma_\mu\right.\right.\nonumber\\                       
&+&\left.\left.\left(a_3 + \bok a_7 + \vq1 a_{11}+ \vq1\bok a_{15}\right)
q_{1\mu} p_{1\nu} \vphantom{\frac{1}{p_1\cdot k}}\right] -\frac{1}{2}\left(a_8
+ a_{16} \vq1\right) \gamma_\mu \gamma_\nu\right\} F^{\mu\nu}. \nonumber  
\end{eqnarray} 
Note that these two forms contain a potential kinematic singularity at $p_1\cdot
k$=0. However, this singularity is outside the physically accessible region for
the process we are discussing, and since this calculation involves no loop
integrations, such singularities are of no real concern.

A further four structures may be eliminated by use of so-called equivalence
relations, leaving a total of eight. A more detailed discussion of this is
beyond the scope of this manuscript.

\section{Phenomenological Lagrangians}

The framework in which we treat the process $\gamma N\to N K\overline{K}$ is the
phenomenological Lagrangian. In this approach, all particles are treated as
point-like. Their structure is accounted for by inclusion of phenomenological
form factors, which we discuss in a later subsection.

\subsection{Ground State Baryons}

We begin with the Lagrangians needed for the electromagnetic vertices of
pseudoscalar mesons and ground state baryons. We treat nucleons as an isospin
doublet, with $N=\left(\matrix{p\cr n}\right)$. Kaons are also treated as 
isospin doublets ($K=\left(\matrix{K^+\cr K^0}\right)$). $\pi$ and $\Sigma$ are
treated as isotriplets.

In what should be a transparent notation, the electromagnetic part of the
Lagrangian is (omitting the $\Theta^+$ for the time being)
\begin{eqnarray}
{\cal L}_1&=&\overline{N}\left(-\frac{e}{2}(1+\tau_3)\gamma_\mu A^\mu+
\frac{e}{4M_N}(k_s^N+\tau_3k_v^N)\gamma_\mu\gamma_\nu F^{\mu\nu}\right)N \nonumber\\
&+&\overline{{\bf \Sigma}}\left(-\frac{e}{2}(1+T_3)
\gamma_\mu A^\mu+\frac{e}{4M_\Sigma}(k_s^\Sigma+\tau_3k_v^\Sigma)
\gamma_\mu\gamma_\nu F^{\mu\nu}\right){\bf \Sigma}\nonumber\\
&+&\overline{\Lambda}\frac{e}{4M_\Lambda}\mu_\Lambda\gamma_\mu
\gamma_\nu F^{\mu\nu}\Lambda
+\overline{\Sigma^0}
\frac{e}{2(M_\Sigma^0+M_\Lambda)}\mu_{\Sigma\Lambda}\gamma_\mu\gamma_\nu 
F^{\mu\nu}\Lambda\nonumber \\ 
&-&\frac{e}{2}\left[K^\dag\left(1+\tau_3\right)\left(\partial_\mu K\right)
-\left(\partial_\mu K^\dag\right)\left(1+\tau_3\right) K\right]A^\mu +H.c.
\end{eqnarray}
where $\mu_{\Sigma\Lambda}$ is the $\Sigma^0\to\Lambda$ transition magnetic 
moment, $\mu_\Lambda$ is the magnetic moment of the $\Lambda$, $k_s^N$ and $k_v^N$ describe the
anomalous magnetic moments of the nucleon doublet, and the $k^\Sigma_{s,v}$ are the corresponding
quantities for the ${\bf \Sigma}$ isotriplet. $T_3$ is the isospin operator for the isotriplet.

The coupling of pseudoscalar mesons to ground state baryons is described by the
Lagrangian
\begin{eqnarray}
{\cal L}_2&=&\frac{g_{NN\pi}}{2M_N}\overline{N}\gamma_\mu\gamma_5 (\partial^\mu 
{\bf \pi\cdot\tau})N
+\frac{g_{N\Lambda K}}{M_N+M_\Lambda}\overline{N}\gamma_\mu\gamma_5 (\partial^\mu K)
\Lambda
+\frac{g_{N\Sigma K}}{M_N+M_\Sigma}\overline{N}\gamma_\mu\gamma_5 {\bf \Sigma\cdot\tau} \partial^\mu
K+ \frac{g_{NN\eta}}{2M_N}\overline{N}\gamma_\mu\gamma_5 N(\partial^\mu 
\eta)\nonumber\\
&-&e\frac{g_{NN\pi}}{2M_N}\overline{N}\gamma_\mu\gamma_5 A^\mu\tau_3 
{\bf \pi\cdot\tau}N
-e\frac{g_{N\Lambda K}}{M_N+M_\Lambda}\overline{N}\gamma_\mu\gamma_5 A^\mu\tau_3 K
\Lambda
-e\frac{g_{N\Sigma K}}{M_N+M_\Sigma}\overline{N}\gamma_\mu\gamma_5 {\bf \Sigma\cdot\tau} A^\mu\tau_3 K+H. c.
\end{eqnarray}
$\eta$ is an isosinglet field representing the $\eta$ meson. The last three terms of this Lagrangian are obtained by minimal substitution in the first three
terms.

\subsection{Vector Mesons}

The vector mesons that enter into our model are $K^*$ and $\phi$. The $K^*$ is
treated as a vector isodoublet field $K_\mu$, completely analogously to the
$K$, while the $\phi$ is represented by a vector isosinglet field $\phi_\mu$.
The Lagrangian in this sector is
\begin{eqnarray}\label{equa98b_sing}
{\cal L}_3&=& \overline{N}\left(G_v^\phi \gamma^\mu \phi_\mu
+i\frac{G_t^\phi}{2 M_N} \gamma^\mu \gamma^\nu\left(\partial_\nu 
\phi_\mu\right)\right)N 
+ \overline{N}\left(G_v^{K^*N\Lambda} \gamma^\mu K^*_\mu
+i\frac{G_t^{K^*N\Lambda}}{M_N+M_\Lambda} \gamma^\mu \gamma^\nu
\left(\partial_\nu K^*_\mu\right)\right)\Lambda\nonumber\\ 
&+& \overline{N}\left(G_v^{K^*N\Sigma} \gamma^\mu {\bf \Sigma\cdot\tau}K^*_\mu
+i\frac{G_t^{K^*N\Sigma}}{M_N+M_\Sigma} \gamma^\mu \gamma^\nu{\bf \Sigma\cdot\tau} \partial_\nu 
K^*_\mu\right)\nonumber\\ 
&+&\epsilon^{\alpha \beta\mu \nu}\left(\frac{g_{\phi \pi \gamma}}{m_\pi}
 \phi_\alpha(\partial_\mu A_\beta)\partial_\nu\pi^0
+\frac{g_{\phi \eta \gamma}}{m_\eta}
 \phi_\alpha(\partial_\mu A_\beta)\partial_\nu\eta\right)
\nonumber\\ 
&+&\frac{g_{\phi KK}}{m_K}
\left[K^\dag\left(\partial^\mu K\right)-\left(\partial^\mu K^\dag\right)
K\right]\phi_\mu
+\frac{g_{K^* K\pi}}{m_K}
\left[K^\dag\left(\partial^\mu{\bf \pi\cdot\tau}\right)-
\left(\partial^\mu K^\dag\right)){\bf \pi\cdot\tau}\right]K^*_\mu.
\end{eqnarray}

\subsection{Baryon Resonances}

There are a number of resonances that need to be taken into account in a
calculation such as this. Since the experimental target is a nucleon, any of
the nucleon or $\Delta$ resonances are expected to play a role. For the energy
range that we consider, and more particularly, for the scope of this
calculation, we find that the most salient points can be illustrated without
any non-strange resonances. Among the hyperons, any number of them can be
included, but again we limit the scope so that only the lowest few hyperon
resonances are taken into account. In either case, we do not consider any
baryon with spin greater than 3/2. With the scope of the model limited like
this, there are only a few Lagrangian terms that must be considered in this
sector. The non-exotic hyperons that are included are listed in table 
\ref{hyperons}.

\subsubsection{Spin 1/2} 

Lagrangian terms needed for spin-1/2 resonances are  
\begin{eqnarray}
{\cal L}_4&=&
\overline{N}\frac{g^{(\frac{1}{2})}_{\Sigma^* NK}}{m_K}\gamma_\mu\gamma_5
{\bf \Sigma^*\cdot\tau}\partial^\mu K 
+ 
\overline{N}\frac{g^{(\frac{1}{2})}_{\Lambda^* NK}}{m_K}\gamma_\mu\gamma_5
\left(\partial^\mu K\right)\Lambda^* 
+ 
\overline{N}\frac{g^{(\frac{1}{2})}_{\Theta NK}}{m_K}\gamma_\mu\gamma_5
\left(\partial^\mu K\right)\Theta_+\nonumber\\
&+&
\overline{N}\frac{g^{(\frac{1}{2})}_{\Sigma^* NK}}{m_K}\gamma_\mu
{\bf \Sigma^*\cdot\tau}\partial^\mu K 
+ 
\overline{N}\frac{g^{(\frac{1}{2})}_{\Lambda^* NK}}{m_K}\gamma_\mu
\left(\partial^\mu K\right)\Lambda^* 
+ 
\overline{N}\frac{g^{(\frac{1}{2})}_{\Theta NK}}{m_K}\gamma_\mu
\left(\partial^\mu K\right)\Theta_- + H.c.,
\end{eqnarray}
where $\Theta_\pm$ is the field for $\Theta^+$ with $J^P=1/2^\pm$. The first three terms of this
Lagrangian correspond to states with $J^P=1/2^+$, while the last three terms are
for $J^P=1/2^-$. In addition, the $\Theta^+$ part of the Lagrangian written
above assumes that the state is an isosinglet. For an isotriplet $\Theta^+$ with
$J^P=1/2^+$, the Lagrangian would become
\begin{equation}
{\cal L}_5=\overline{N}\frac{g_{\Theta NK}}{m_K}\gamma_\mu\gamma_5
{\bf \Theta_+\cdot\tau}\partial^\mu K+
\overline{N}\frac{g_{\Theta NK}}{m_K}\gamma_\mu
{\bf \Theta_-\cdot\tau}\partial^\mu K.
\end{equation}

\subsubsection{Spin 3/2}

The Lagrangian terms for spin-3/2 resonances are
\begin{eqnarray}\label{equa101b_sing}
{\cal L}_6&=&
\overline{N}\frac{g^{(\frac{3}{2})}_{\Sigma^* NK}}{m_K}
{\bf \Sigma^*_\mu\cdot\tau}\partial^\mu K 
+ 
\overline{N}\frac{g^{(\frac{3}{2})}_{\Lambda^* NK}}{m_K}
\left(\partial^\mu K\right)\Lambda^*_\mu 
+ 
\overline{N}\frac{g^{(\frac{3}{2})}_{\Theta NK}}{m_K}
\left(\partial^\mu K\right)\Theta_{+\mu}\nonumber\\
&+&
\overline{N}\frac{g^{(\frac{3}{2})}_{\Sigma^* NK}}{m_K}\gamma_5
{\bf \Sigma^*_\mu\cdot\tau}\partial^\mu K 
+ 
\overline{N}\frac{g^{(\frac{3}{2})}_{\Lambda^* NK}}{m_K}\gamma_5
\left(\partial^\mu K\right)\Lambda^*_\mu 
+ 
\overline{N}\frac{g^{(\frac{3}{2})}_{\Theta NK}}{m_K}\gamma_5
\left(\partial^\mu K\right)\Theta_{-\mu} + H.c.,
\end{eqnarray}
where the $\mu$ indices on the $\Lambda$, $\Sigma$ and $\Theta$ fields indicate
that they are vector-spinor, spin-3/2 fields. In this calculation, we use the
Rarita-Schwinger version of such fields. The first three terms are for resonances
with positive parity, while the last three are for resonances with negative
parity. For an isovector $\Theta$ the 
Lagrangian terms are
\begin{equation}
{\cal L}_7=\overline{N}\frac{g_{\Theta NK}}{m_K}
{\bf \Theta_{+\mu}\cdot\tau}\partial^\mu K+i\overline{N}\frac{g_{\Theta NK}}{m_K}
\gamma_5{\bf \Theta_{-\mu}\cdot\tau}\partial^\mu K,
\end{equation}
where the $\Theta_{\pm \mu}$ represents a $\Theta^+$ with $J^P=3/2^\pm$,
respectively.

\subsection{Coupling Constants}

To evaluate the coupling constants of the ground-state baryons to pseudoscalar mesons, we use the
extended Goldberger-Treimann relations. For the coupling of the baryons $B$ and
$B^\prime$ to the pseudoscalar $M$, the relation is
\begin{equation}
g_{BB^\prime M}=\left(\frac{G_A}{G_V}\right)_{B\to B^\prime} \frac{M_B+M_{B^\prime}}{2f_M},
\end{equation}
where $f_M$ is the meson decay constant for the pseudoscalar meson $M$. 
$\left(\frac{G_A}{G_V}\right)_{B\to B^\prime}$ is obtained from the semileptonic decay of
$B\to B^\prime$ or $B^\prime\to B$. The values of $f_M$, 
$\left(\frac{G_A}{G_V}\right)_{B\to B^\prime}$ (taken from the Review of
Particle Physics \cite{pdg}) and $g_{BB^\prime M}$ obtained
from these relations are shown in table \ref{bornconstants}.

\begin{table}
\caption{Values of $g_{BB^\prime M}$ obtained using the Goldberger-Treimann relations.
\label{bornconstants}}
\begin{tabular}{c|c|c|c}
Coupling & $f_M$ (GeV) & $\left(\frac{G_A}{G_V}\right)_{B\to B^\prime}$ 
& $g_{BB^\prime M}$ \\ \hline
$g_{\pi NN\pi}$ & $\frac{0.13}{\sqrt{2}}$ & 1.22 & 12.8 \\ \hline
$g_{N\Sigma K}$ & $\frac{0.16}{\sqrt{2}}$ & 0.34 & 3.2 \\ \hline
$g_{N\Lambda K}$ & $\frac{0.16}{\sqrt{2}}$ & -0.718 & -6.51 \\ \hline
$g_{NN\eta}$ & $\approx 1.2f_\pi$ & 1.22 & 10.37 \\ 
\end{tabular}
\end{table}

The decay width of a vector meson into two pseudoscalars is related to the corresponding coupling
constant by
\begin{equation}
\Gamma_{V\to P_1P_2}=\frac{g_{VP_1P_2}^2}{48\pi M_V^5}\lambda^\frac{3}{2}
\left(M_V^2,M_{P_1}^2,M_{P_2}^2\right),
\end{equation}
where $\lambda(a,b,c)$ is the K$\not\!a$llen function $\lambda(a,b,c)=a^2+b^2+c^2-2(ab+ac+bc)$. From the
measured widths and branching fractions of the $\phi$ and $K^*$ mesons, we find that 
\begin{equation}
g_{\phi KK}=4.3 ,\,\,\,\, g_{K^*K\pi}=5.6.
\end{equation}
In a similar way, the width for the process $V\to P\gamma$ is
\begin{equation}
\Gamma_{V\to P\gamma}=\frac{g_{VP\gamma}^2}{192\pi M_V^3}\lambda^\frac{3}{2}
\left(M_V^2,M_P^2,0\right),
\end{equation}
which leads to
\begin{equation}
g_{\phi\eta\gamma}=4.3,\,\,\,\, g_{\phi\pi^0\gamma}=0.055,\,\,\,\, 
g_{K^{0*}K^0\gamma}=0.35,\,\,\,\,g_{K^{+*}K^+\gamma}=0.22.
\end{equation}

For a baryon $B$ with $J^P=1/2^+$, the width for the decay into a pseudoscalar 
meson $P$ and a ground-state baryon $B^\prime$ is
\begin{equation}\label{halfp}
\Gamma_{B\to B^\prime P}=\frac{g_{BB^\prime P}^2}{16\pi M_B^3 M_P^2}
\left(M_B+M_{B^\prime}\right)^2 \left[\left(M_B-M_{B^\prime}\right)^2-M_P^2\right]
\lambda^\frac{1}{2}\left(M_B^2,M_{B^\prime}^2,M_P^2\right),
\end{equation}
while the corresponding width for a baryon with $J^P=1/2^-$ is
\begin{equation}
\Gamma_{B\to B^\prime P}=\frac{g_{BB^\prime P}^2}{16\pi M_B^3 M_P^2}
\left(M_B-M_{B^\prime}\right)^2 \left[\left(M_B+M_{B^\prime}\right)^2-M_P^2\right]
\lambda^\frac{1}{2}\left(M_B^2,M_{B^\prime}^2,M_P^2\right).
\end{equation}
For baryons with $J^P=3/2^\pm$, the widths are
\begin{eqnarray}\label{thalf}
\Gamma_{B\to B^\prime P}&=&\frac{g_{BB^\prime P}^2}{192\pi M_B^5 M_P^2}
\left[\left(M_B+M_{B^\prime}\right)^2-M_P^2\right]
\lambda^\frac{3}{2}\left(M_B^2,M_{B^\prime}^2,M_P^2\right),\nonumber\\
\Gamma_{B\to B^\prime P}&=&\frac{g_{BB^\prime P}^2}{192\pi M_B^5 M_P^2}
\frac{\lambda^\frac{5}{2}\left(M_B^2,M_{B^\prime}^2,M_P^2\right)}
{\left[\left(M_B+M_{B^\prime}\right)^2-M_P^2\right]},
\end{eqnarray}
where the first expression is for a positive parity parent baryon. The non-exotic hyperons that are
used in this calculation, along with their masses, total widths, spins, parities, and their $NK$ branching fractions
and coupling constants, obtained from eqns. (\ref{halfp}) - (\ref{thalf}), are shown in table \ref{hyperons}.

\begin{table}[h!]\caption{Values of $g_{YNK}$ for non-exotic hyperons appearing in the model.
\label{hyperons}}
\begin{tabular}{c|l|c|c|c}
$Y$(Mass)& $J^P$  & $\Gamma$ (MeV) & $\frac{\Gamma_{NK}}{\Gamma}$ & $g_{YNK}$ \\ \hline
$\Lambda(1520)$ & $\frac{3}{2}^-$ & 16 & 0.45 & 15.2 \\\hline
$\Lambda(1600)$ & $\frac{1}{2}^+$ & 150 & 0.2 & 1.05 \\\hline
$\Lambda(1670)$ & $\frac{1}{2}^-$ & 35 & 0.25 & 0.32 \\\hline
$\Lambda(1690)$ & $\frac{3}{2}^-$ & 60 & 0.25 & 5.53 \\\hline
$\Lambda(1800)$ & $\frac{1}{2}^+$ & 300 & 0.35 & 0.86 \\\hline
$\Lambda(1810)$ & $\frac{1}{2}^+$ & 150 & 0.35 & 0.71 \\\hline
$\Lambda(1890)$ & $\frac{3}{2}^+$ & 100 & 0.3 & 1.09 \\\hline
$\Sigma(1580)$ & $\frac{3}{2}^-$ & 15 & 0.45 & 1.95 \\\hline
$\Sigma(1620)$ & $\frac{1}{2}^-$ & 80 & 0.22 & 0.52 \\\hline
$\Sigma(1660)$ & $\frac{1}{2}^+$ & 100 & 0.2 & 0.67 \\\hline
$\Sigma(1670)$ & $\frac{3}{2}^-$ & 60 & 0.1 & 3.88 \\\hline
$\Sigma(1750)$ & $\frac{1}{2}^-$ & 90 & 0.26 & 0.44 \\\hline
$\Sigma(1880)$ & $\frac{1}{2}^+$ & 80 & 0.06 & 0.19 \\\hline
$\Sigma(1940)$ & $\frac{3}{2}^-$ & 220 & 0.13 & 3.19 \\
\end{tabular}
\end{table}

As mentioned above, we allow the $\Theta$ pentaquarks to have four different
combinations of spin and parity in
this calculation. In addition, since the width of this particle has not yet been ascertained, we also
allow different widths. Table \ref{thetacouplings} shows the values of the coupling constants
we obtain for the different spins, parities and total widths of the $\Theta$, assuming that the $NK$
final state saturates its decays.
\begin{table}\caption{Values of $g_{\Theta NK}$ for different spins, parities and total widths of the
$\Theta$.
\label{thetacouplings}}
\begin{tabular}{c|l|c|c}
$\hphantom{klk}J\hphantom{klk}$ & $P$ & $\Gamma$ (MeV) & $g_{\Theta NK}$ \\ \hline
$\frac{1}{2}$ & $+$ & 1 & 0.27 \\ \hline
$\frac{1}{2}$ & $+$ & 10 & 0.87 \\ \hline
$\frac{1}{2}$ & $-$ & 1 & 0.16 \\ \hline
$\frac{1}{2}$ & $-$ & 10 & 0.50 \\ \hline
$\frac{3}{2}$ & $+$ & 1 & 0.61 \\ \hline
$\frac{3}{2}$ & $+$ & 10 & 1.94 \\ \hline
$\frac{3}{2}$ & $-$ & 1 & 4.35 \\ \hline
$\frac{3}{2}$ & $-$ & 10 & 13.76 \\
\end{tabular}
\end{table}

Finally, we note that there are a number of coupling constants for which little
information is available. Perhaps the most important of these in terms of
contributions to the cross sections are the couplings of the vector mesons
$K^*$ and $\phi$, particularly those of the $\phi$ to the ground-state baryons.
The couplings of the two hyperon resonances that lie below the $NK$ threshold,
namely the $\Sigma(1385)$ and the $\Lambda(1405)$ are also not known with much
certainty. When we display and discuss our results, we will comment further on
the effects that these coupling constants have on the graphs that we show.

\subsection{Diagrams}

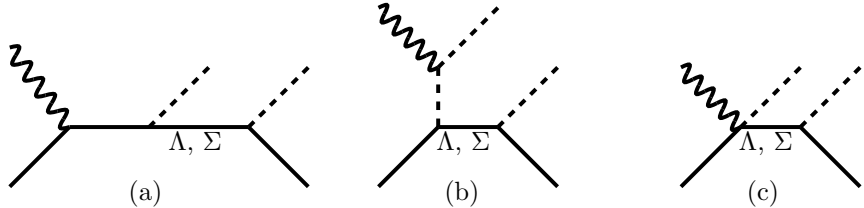
\begin{figure}[h!]
\begin{center}
\begin{picture}(100,50)
\Line(0,0)(15,15)1
\Line(15,15)(60,15)1
\Line(60,15)(75,0)1
\Photon(0,35)(15,15)251
\DashLine(60,15)(75,30)21
\DashLine(35,15)(50,30)21
\put(60,14){$\Lambda$, $\Sigma$}
\put(45,-5){(a)}
\end{picture}
\hskip .5in
 \begin{picture}(75,75)
\Line(0,0)(15,15)1
\Line(15,15)(30,15)1
\Line(30,15)(45,0)1
\Photon(0,45)(15,30)251
\DashLine(15,15)(15,30)21
\DashLine(15,30)(30,45)21
\DashLine(30,15)(45,30)21
\put(22,14){$\Lambda$, $\Sigma$}
\put(25,-5){(b)}
\end{picture}
\hskip .5in
 \begin{picture}(75,50)
\Line(0,0)(15,15)1
\Line(15,15)(30,15)1
\Line(30,15)(45,0)1
\Photon(0,30)(15,15)251
\DashLine(15,15)(30,30)21
\DashLine(30,15)(45,30)21
\put(22,14){$\Lambda$, $\Sigma$}
\put(25,-5){(c)}
\end{picture}
\end{center}

\caption{`Born' diagrams: continuous, unlabeled lines are nucleons. Unlabeled dashed lines
are kaons and wavy lines are photons.\label{fig2}}
\end{figure}

\vskip 0.5in

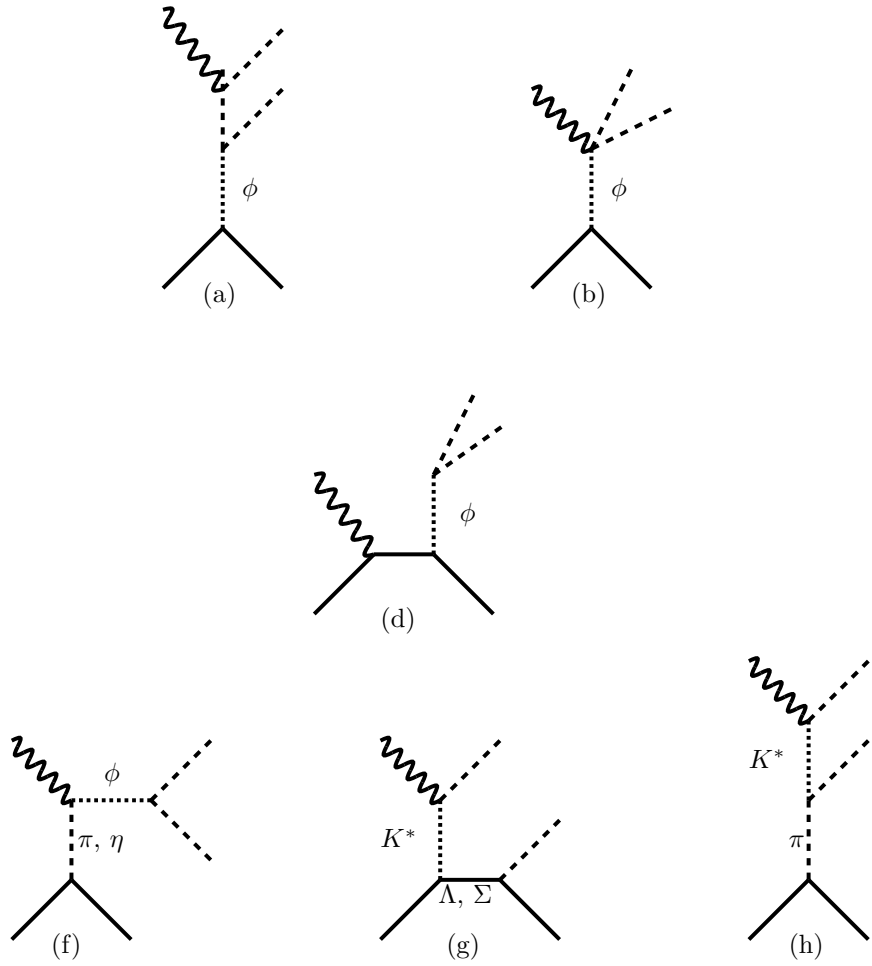
\begin{figure}
\begin{center}

\begin{picture}(100,50)
\Line(0,0)(15,15)1
\Line(15,15)(30,0)1
\DashLine(15,15)(15,35)11
\Photon(0,70)(15,50)251
\DashLine(15,35)(15,55)21
\DashLine(15,35)(30,50)21
\DashLine(15,50)(30,65)21
\put(15,-5){(a)}
\put(30,35){$\phi$}
\end{picture}
\hskip .5in
 \begin{picture}(75,75)
\Line(0,0)(15,15)1
\Line(15,15)(30,0)1
\DashLine(15,15)(15,35)11
\DashLine(15,35)(35,45)21
\DashLine(15,35)(25,55)21
\Photon(0,50)(15,35)251
\put(15,-5){(b)}
\put(30,35){$\phi$}
\end{picture}
\vskip 1.in
\begin{picture}(100,50)
\Line(0,0)(15,15)1
\Photon(0,35)(15,15)251
\Line(15,15)(30,15)1
\Line(30,15)(45,0)1
\DashLine(30,15)(30,35)11
\DashLine(30,35)(40,55)21
\DashLine(30,35)(47,47)21
\put(55,35){$\phi$}
\put(25,-5){(d)}
\end{picture}
\vskip 1.in
\begin{picture}(100,50)
\Line(0,0)(15,15)1
\DashLine(15,15)(15,35)21
\Line(15,15)(30,0)1
\Photon(0,50)(15,35)251
\DashLine(15,35)(35,35)11
\DashLine(35,35)(50,50)21
\DashLine(35,35)(50,20)21
\put(35,60){$\phi$}
\put(25,35){$\pi$, $\eta$}
\put(15,-5){(f)}
\end{picture}
\hskip 0.5in
\begin{picture}(100,50)
\Line(0,0)(15,15)1
\DashLine(15,15)(15,35)11
\Photon(0,50)(15,35)251
\DashLine(15,35)(30,50)21
\Line(15,15)(30,15)1
\Line(30,15)(45,0)1
\DashLine(30,15)(45,30)21
\put(25,-5){(g)}
\put(0,35){$K^*$}
\put(22,14){$\Lambda$, $\Sigma$}
\end{picture}
\hskip 0.5in
\begin{picture}(50,50)
\Line(0,0)(15,15)1
\Line(15,15)(30,0)1
\DashLine(15,15)(15,35)21
\DashLine(15,35)(15,55)11
\DashLine(15,35)(30,50)21
\Photon(0,70)(15,55)251
\DashLine(15,55)(30,70)21
\put(15,-5){(h)}
\put(0,65){$K^*$}
\put(15,35){$\pi$}

\end{picture}
\vskip .5in
\caption{ Diagrams containing vector mesons. The dotted lines represent 
the vector mesons.\label{fig3}}
\end{center}
\end{figure}

\begin{figure}
\begin{center}

\hskip .4in
\begin{picture}(100,50)
\Line(0,0)(15,15)1
\Line(15,15)({37.5},15)1
\Line({37.5},15)(60,15){2.5}
\Line(60,15)(75,0)1
\Photon(0,35)(15,15)251
\DashLine(60,15)(75,30)21
\DashLine(35,15)(50,30)21
%\put(60,5){$\matrix{\Lambda^*,\Sigma^*\cr\Theta}$}
\put(45,-5){(a)}
\end{picture}
\hskip 1.25in
 \begin{picture}(50,50)
\Line(0,0)(15,15)1
\Line(15,15)(30,15){2.5}
\Line(30,15)(45,0)1
\Photon(0,45)(15,30)251
\DashLine(15,15)(15,30)21
\DashLine(15,30)(30,45)21
\DashLine(30,15)(45,30)21
\put(25,-5){(b)}
\end{picture}
\vskip 1in
 \begin{picture}(75,0)
\Line(0,0)(15,15)1
\Line(15,15)(30,15){2.5}
\Line(30,15)(45,0)1
\Photon(0,30)(15,15)251
\DashLine(15,15)(30,30)21
\DashLine(30,15)(45,30)21
\put(35,-10){(c)}
\end{picture}
\hskip 1.25in
\begin{picture}(50,0)
\Line(0,0)(15,15)1
\Line(15,15)({37.5},15){2.5}
\Line({37.5},15)(60,15){2.5}
\Line(60,15)(75,0)1
\Photon(25,35)(40,15)251
\DashLine(60,15)(75,30)21
\DashLine(15,15)(30,30)21
\put(45,-5){(d)}
\end{picture}
\end{center} 

\caption{Diagrams containing excited baryons. In (a) to (d), the
thick solid lines may be either $\Lambda^*$, $\Sigma^*$ or $\Theta$, while the
thin solid line is a nucleon. In diagram (d), the photon couples to the charge
of the intermediate resonance: in this model, we neglect couplings to any higher
moments of the resonance.\label{fig4}} 
\end{figure}
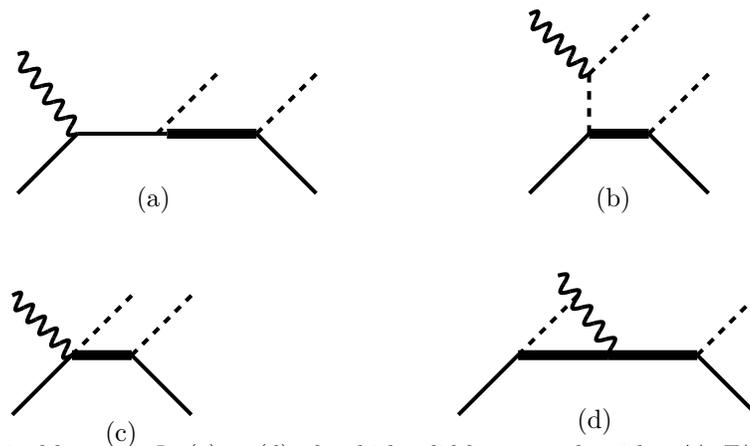

The diagrams that we include in this calculation are shown in figures
\ref{fig2} to \ref{fig4}. In these diagrams, solid lines represent baryons. If
a solid line is unlabeled, it represents a nucleon. Dashed lines represent
pseudoscalar mesons, with unlabeled dashed lines representing kaons. Wavy
lines are photons, and dotted lines are vector mesons. Each diagram shown
actually represents a set of diagrams, as all allowed permutations of external
meson and photon legs are taken into account.

We include a number of hyperon resonances in this calculation. These are listed in table
\ref{hyperons}. For each of the resonances, there is a corresponding set of diagrams of the kind
shown in figure \ref{fig4}.

There are a number of contributions that have been omitted from this calculation. For instance,
we have omitted all but the ground-state nucleon, and all of the $\Delta$ resonances. In fact,
with the information that is available on how these states couple to final states with hidden
strangeness, we have found that their contributions to the cross section are small. We have also
neglected couplings to higher moments of any of the hyperon resonances (figure \ref{fig4} (d)),
as well as any contributions that would arise from electromagnetic transitions between excited
hyperons and their ground states. In principle, there is no {\it a priori} reason to expect such
contributions to be small, but little is known of those couplings. Including such contributions
would add too many unknown parameters to the model.

\subsection{Form Factors}

Apart from the photon, none of the states that enter this calculation are
elementary particles: they all have substructure, and this substructure is
reflected in the need to include some kind of form factor at each interaction
vertex. Indeed, without such form factors, cross sections grow with energy, and
the unitarity limit is quickly violated.

Inclusion of any form factors in a calculation like this must be done in a
manner that preserves gauge invariance, and a detailed discussion of all of the
issues that arise, and all of the methods and prescriptions for preserving gauge
invariance, are beyond the scope of this manuscript. In this calculation, we
adopt the prescription of assigning an overall form factor to gauge invariant
sets of diagrams. This means, for instance, that all of the diagrams of the type
shown in fig. \ref{fig2} have the same form factor as a multiplicative factor.
For all of the form factors, we choose the form \cite{nam,oh}
\begin{equation}
F=\left(\frac{X^4}{(p_i^2-m_i^2)^2+X^4}\right)^n.
\end{equation}

In this expression, $p_i$ is usually the momentum of the off-shell particle
with mass $m_i$. In this calculation, we make the simplification of setting all
of the $p_i^2$ to be equal to $s$, the total energy in the cm frame, squared.
$X$ is chosen to be 1.8 GeV, as  has been used by other authors. In addition,
since we apply this form factor to sets of diagrams, we choose $m_i$ to be the
mass of the lightest  off-shell particle in a particular set. The exception
to this occurs in the diagrams of fig. \ref{fig3} (f) and (h), where $m_i$ is
chosen to be the mass of the vector meson in the diagram. The value of the integer $n$ depends on
the spin of baryons in the diagram. If there are only spin-1/2 baryons in the
set of diagrams, $n$ is chosen to be unity, while for spin-3/2 baryons, $n$ is
chosen to be two.

The form that we have chosen for the form factors, as well as the manner in
which we apply them, is simply a prescription, and is not meant to be rigorous.
We note that the form factors chosen have the desired effect, producing cross
sections that are roughly of the correct order of magnitude. Without these form
factors, calculated cross sections are much too large.

\section{Results}

There are six possible channels to be explored, namely $\gamma p\to pK^+K^-$,
$\gamma p\to pK^0\overline{K}^0$, $\gamma p\to nK^+\overline{K}^0$,
$\gamma n\to nK^+K^-$, $\gamma n\to nK^0\overline{K}^0$ and 
$\gamma n\to pK^0K^-$. It will be impossible to present results for all of these
channels without making this manuscript overly long. We therefore choose a few
examples to illustrate the main features of the model. We note, however, that
the first result reported from JLab used a deuteron target, while searches using
proton targets have been and are being carried out. Examining the cross sections
for both kinds of targets is therefore relevant.

In the following subsections, we present the results of our model calculation.
We begin with the results of the full model, including the $\phi$ and
$\Lambda(1520)$. We then exclude these two states to more closely simulate the
pentaquark searches that have been carried out at JLab, and examine the effects
of the spin, parity, width and isospin of the pentaquark on the cross section.
We also examine the role of the $K^*$ in increasing the cross section for
production of the $\Theta^+$.

We note that the $\Sigma(1385)$ plays very little role in the results we
present, as its contribution to the cross section is small. The same is true of
the $N^*$ states that we consider, as their couplings to $\Lambda K$ and
$\Sigma K$ final states are generally small, at least for the ones we have 
examined. The $\Lambda(1405)$, on the other hand, can significantly affect the
cross section near threshold in the $N\overline{K}$ mass distributions. If the
value  for the coupling of this state to the $N\overline{K}$ channel is chosen
to be sufficiently large, a sharp shoulder at lower masses arises in the mass
distribution of the $N\overline{K}$. The absence of such a shoulder in the
experimental data limits the size of this coupling constant. We use a value of
5.3 for this constant. 

In reference \cite{clas}, the process studied was $\gamma d\to pnK^+K^-$. The
$\Lambda(1520)$ was identified in the mass distribution of the $pK^-$ pair,
the $\phi$ in the $K^+K^-$ pair, and the $\Theta^+$ in the $nK^+$ pair. We
assume that either one of the initial nucleons takes an active part in the
scattering process, while the other acts as a spectator. This would mean that
the two processes contributing to the $pnK^+K^-$ final state are $\gamma p\to
pK^+K^-$ and $\gamma n\to nK^+K^-$. In addition, this means that in the mass
distributions observed, only the proton component of the target contributes to
the production of the $\Lambda(1520)$, and only the neutron component
contributes to the production of the isoscalar $\Theta^+$. The two processes
$\gamma p\to pK^+K^-$ and  $\gamma n\to nK^+K^-$ are therefore the focus of
much of our discussion. However, searches in other channels have been and are
being carried out, and some discussion is devoted to those channels as well.

\subsection{Full Model}

For all of the results that we display, we present 
$\partial\sigma/\partial m_{ij}^2$, where $m_{ij}^2=(p_i+p_j)^2$, and $p_i$ is the momentum of the $i$th particle in
the final state. Thus, we expect to see strong resonant effects from the $\phi$
in the $K\overline{K}$ subsystem, and similar effects arising from the
$\Lambda(1520)$ in the $N\overline{K}$ subsystem. We also expect to see weaker resonant effects from
the other hyperons that are included in the calculation.

While the coupling of the $\Lambda(1520)$ to $NK$ can be determined from the
$NK$ partial width of the state, there is no simple way of determining the $\phi
NN$ coupling constants, except by a detailed analysis of $\phi$
photoproduction cross sections. Indeed, the angular distributions would have to
be analyzed in order to determine the relative magnitudes and signs of the
vector and tensor couplings. Such an analysis is well beyond the scope of this
work. In the results that we show for the full calculation, we choose
$G_v^\phi=4$ and $G_t^\phi=5$. The actual values are of no import for the main
topic of the article. We must also point out that we have not included any diffractive
production of the $\phi$.

The results that we show are for $w=$ 2.5 GeV. We have examined some of the
cross sections for smaller values of $w$, and will comment on those results
later in the article.

\subsubsection{Isoscalar, Spin-1/2 $\Theta^+$}

\begin{figure}
\centerline{\includegraphics[width=3in,angle=-90]{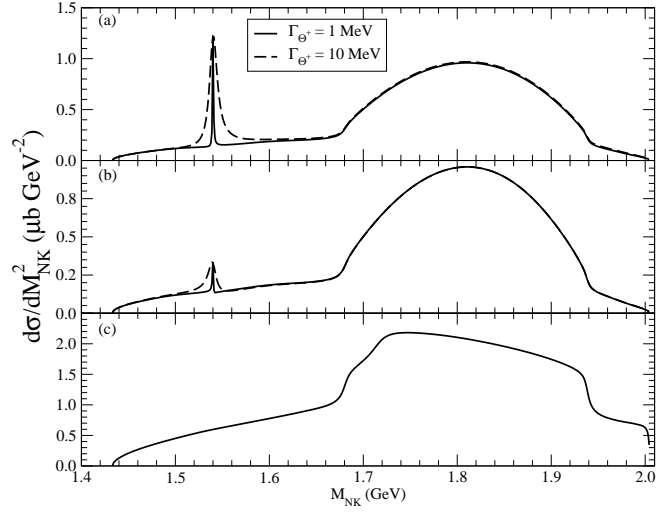}}
\caption{The differential cross section $\partial\sigma/\partial
M_{NK}^2$ as a function of $M_{NK}$, for a spin-1/2 $\Theta^+$. The curves in
(a) and (b) are for the process  $\gamma n\to nK^+K^-$, while the curve in (c) 
is for $\gamma p\to pK^+K^-$. In (a) and (b), the solid curves arise for a
pentaquark with a width of 1 MeV, while the dashed curves correspond to a width of
10 MeV. The curves in (a) arise from a pentaquark with positive parity, while
those in (b) are for a pentaquark of negative parity.\label{set1a}} \end{figure}

Figure \ref{set1a} shows the differential cross section,
$\partial\sigma/\partial M_{NK}^2$, for an isoscalar $\Theta^+$ with spin 1/2.
The curves in (a) and (b) are for the process $\gamma n\to nK^+K^-$, while (c)
is for $\gamma p\to pK^+K^-$. In (a) and (b),  the width of the $\Theta^+$ is
allowed to be 1 MeV (solid curves) or 10 MeV (dashed curves). In addition, (a)
results from a $\Theta^+$ with positive parity, while (b) corresponds to one
with negative parity. Since the isoscalar $\Theta^+$ is not produced off the
proton in this channel, neither its parity nor its width affects the curve that
results in (c). The structures seen in this plot arise from kinematic
reflections of the $\Lambda(1520)$ and the $\phi$.

If there were free neutron targets, the curves in (a) suggest that the
$\Theta^+$ would be relatively easy to observe above background, modulo
detector efficiency, resolution and acceptance issues. However, for deuteron targets, the
presence of the proton would modify this somewhat. 

\begin{figure}
\centerline{\includegraphics[width=3in,angle=-90]{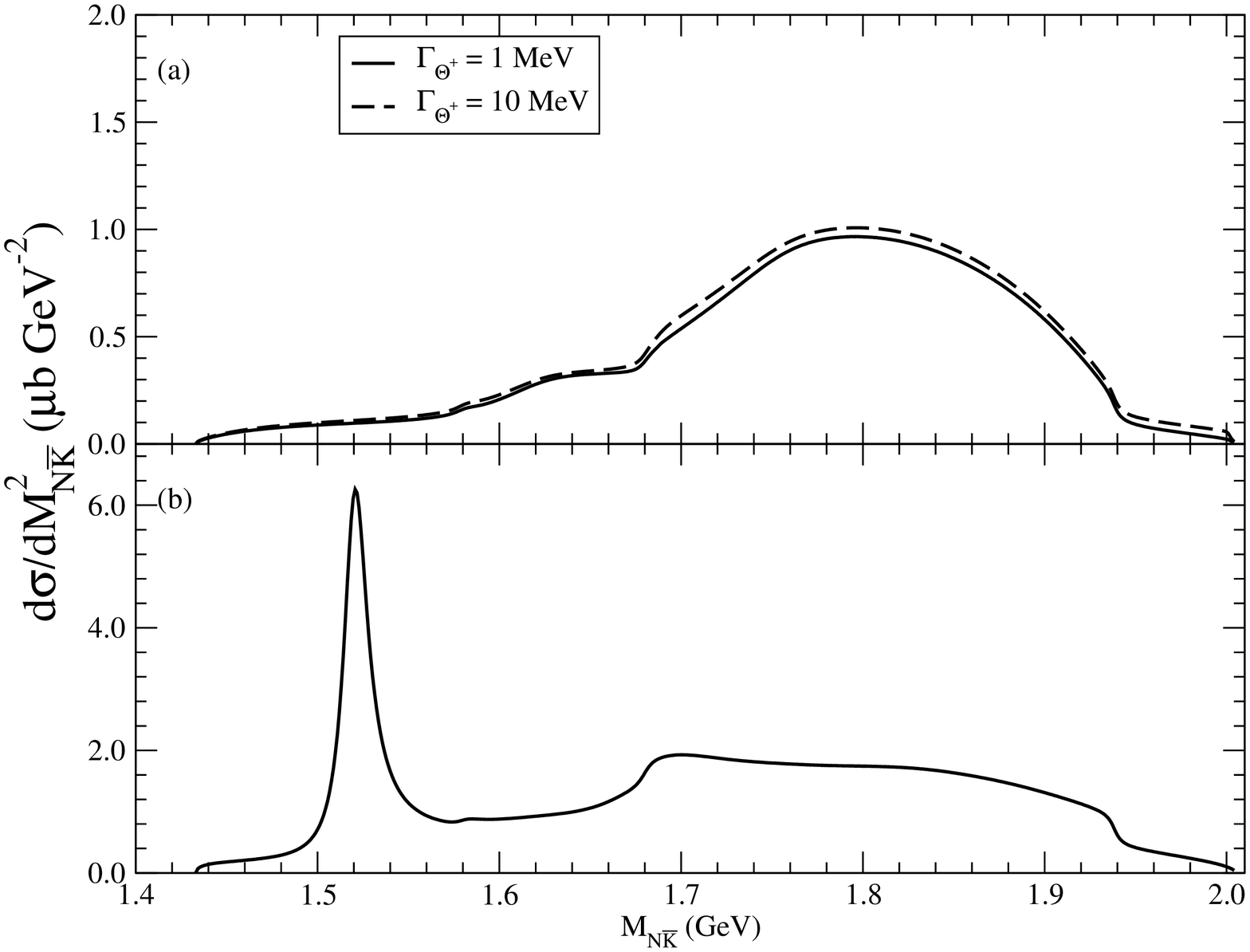}\hspace*{-0.2in}
\includegraphics[width=3in,angle=-90]{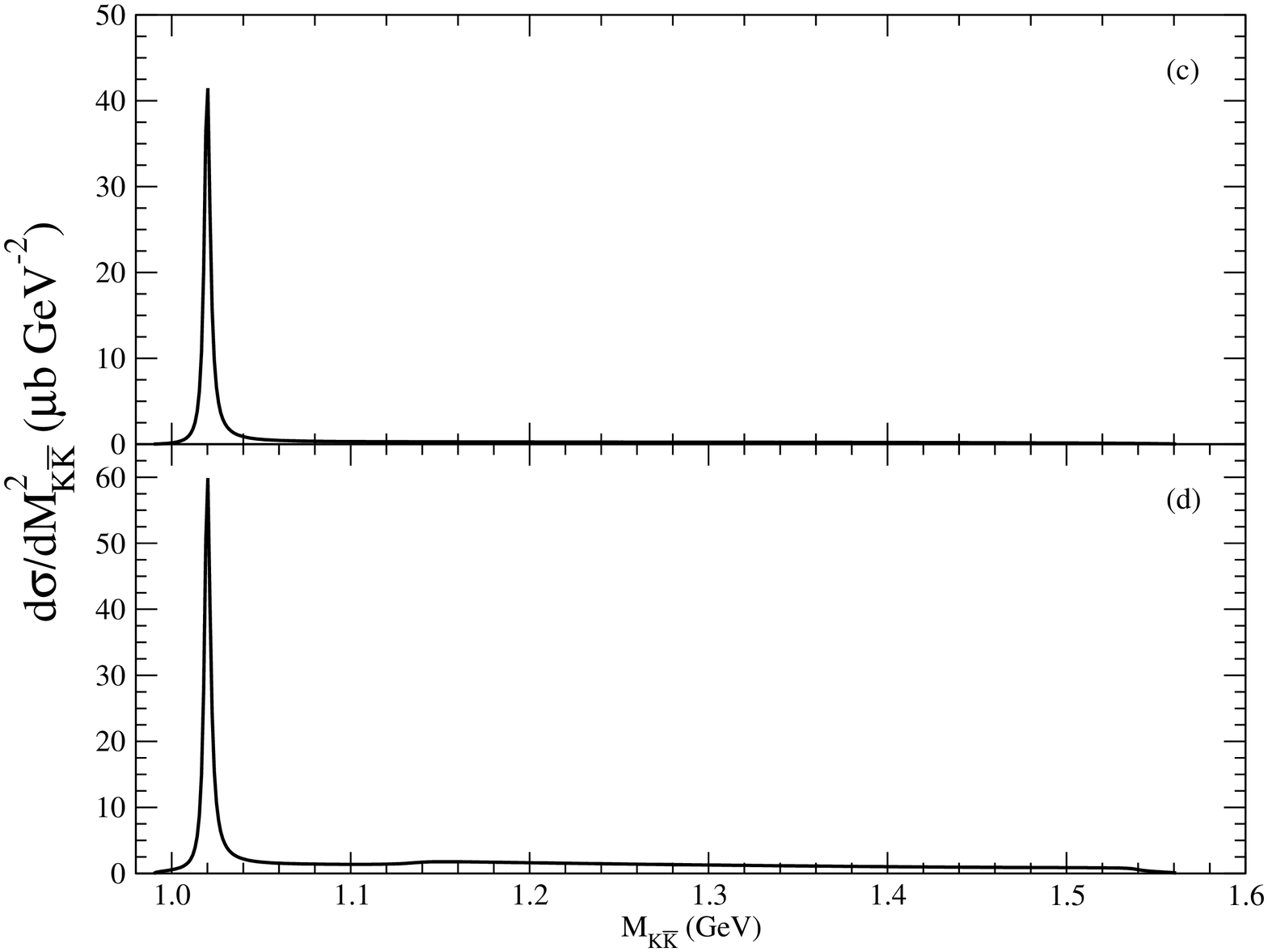}}
\caption{The differential cross section $\partial\sigma/
\partial M_{N\overline{K}}^2$ as a function of $M_{N\overline{K}}$ ((a) and 
(b)), and $\partial\sigma/\partial M_{K\overline{K}}^2$ as a function of 
$M_{K\overline{K}}$ ((c) and (d)). The curves in (a) and (c) are for the 
process $\gamma n\to nK^+K^-$, while those in (b) and (d) are for 
$\gamma p\to pK^+K^-$.\label{set1bc}}
\end{figure}

Figure \ref{set1bc} shows the same differential cross sections, but as
functions of different invariant masses. Figures \ref{set1bc} (a) and (b) show
the differential cross sections as a function of the mass of the
nucleon-antikaon pair, while (c) and (d) show it as a function of the mass of
the $K\overline{K}$ pair. In addition, (a) and (c) result from a neutron
target, while (b) and (d) are from a proton. In the case of the proton target, 
the $\Lambda(1520)$ dominates the cross section in (b), while the contribution
of the $\phi$ can be seen as the structure at larger values of $M_{N\overline{K}}$.
The roles are reversed in (d): the $\phi$ gives the prominent peak, while the
$\Lambda(1520)$ provides the `plateau' at larger invariant mass. For the
neutron target ((a) and (c)), $\Lambda$'s do not contribute to this channel,
but the effects of the $\Sigma$ resonances included in the calculation can be
seen in (a). In this case, the bulk of the cross section comes from the $\phi$,
as can be seen from (c).

\begin{figure}
\centerline{\includegraphics[width=3in,angle=-90]{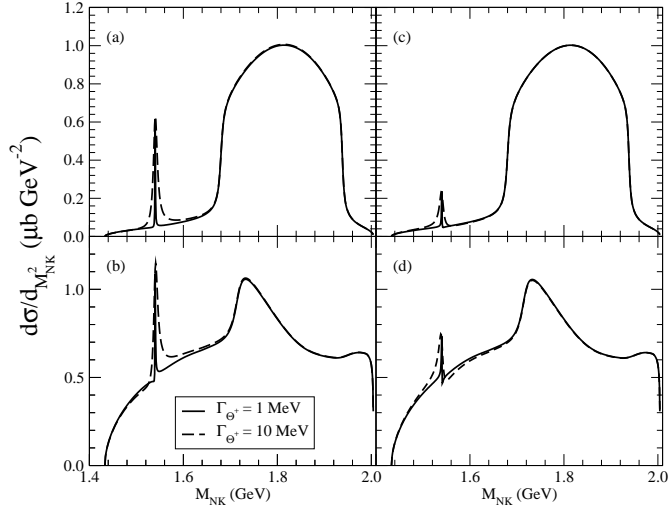}}
\caption{The differential cross section $\partial\sigma/\partial
M_{NK}^2$ as a function of $M_{NK}$. The curves in (a) and (c) are for the
process $\gamma p\to pK^0\overline{K}^0$, while those in (b) and (d) are for
$\gamma p\to nK^+\overline{K}^0$. (a) and (b) are for a positive parity
$\Theta^+$ while (c) and (d) are for a $\Theta^+$ with negative
parity.\label{set1d}} \end{figure}

Figure \ref{set1d} shows the differential cross section,
$\partial\sigma/\partial M_{NK}^2$, for the processes $\gamma p\to p
K^0\overline{K}^0$ ((a) and (c)) and $\gamma p\to n K^+\overline{K}^0$ ((b) and
(d)). The graphs in (a) and (b) assume that the $\Theta^+$ has positive parity,
while those in (c) and (d) are for a pentaquark with negative parity. From
these curves, particularly those in (b) and (d), it should be clear that
observing a $\Theta^+$ signal could be somewhat problematic unless the
contributions from the $\Lambda(1520)$ and the $\phi$ were excluded. We discuss
this in a later subsection.

\subsubsection{Isoscalar, Spin-3/2 $\Theta^+$}

\begin{figure}
\centerline{\includegraphics[width=3in,angle=-90]{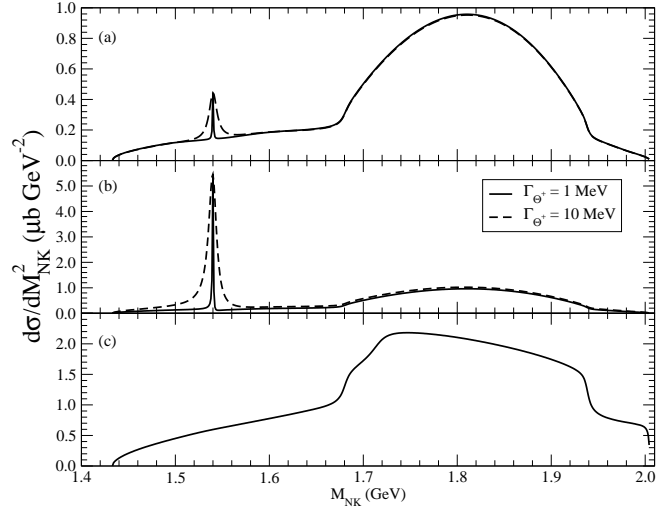}}
\caption{The differential cross section $\frac{\partial\sigma}{\partial
M_{NK}^2}$ as a function of $M_{NK}$, for a spin-3/2 $\Theta^+$. The curves in
(a) and (b) are for the process $\gamma n\to nK^+K^-$, while (c) is
for $\gamma p\to pK^+K^-$. In (a), the $\Theta^+$ has positive parity, while in
(b), its parity is negative.\label{set2a}} \end{figure}

Figure \ref{set2a} shows the differential cross section if the $\Theta^+$ is
assumed to be an isoscalar with spin 3/2. As with the spin-1/2 discussion, the
curves in (a) and (b) are for the process $\gamma n\to nK^+K^-$, while those in
(c) are for $\gamma p\to pK^+K^-$. In (a) and (b), the width of the $\Theta^+$
is allowed to be 1 MeV (solid curves) or 10 MeV (dashed curves). In addition,
(a) results from a $\Theta^+$ with positive parity, while (b) corresponds to
one with negative parity. It is interesting to note that the height of the peak of the
$\Theta^+$ for the $3/2^-$ case is comparable to that of the $\Lambda(1520)$
(seen in figure \ref{set1bc} (b), for instance).
This is not surprising, since the states are almost degenerate, and the height
of the peak at resonance depends only on kinematics, which would be largely the
same for the two resonances.

\begin{figure}
\centerline{\includegraphics[width=3in,angle=-90]{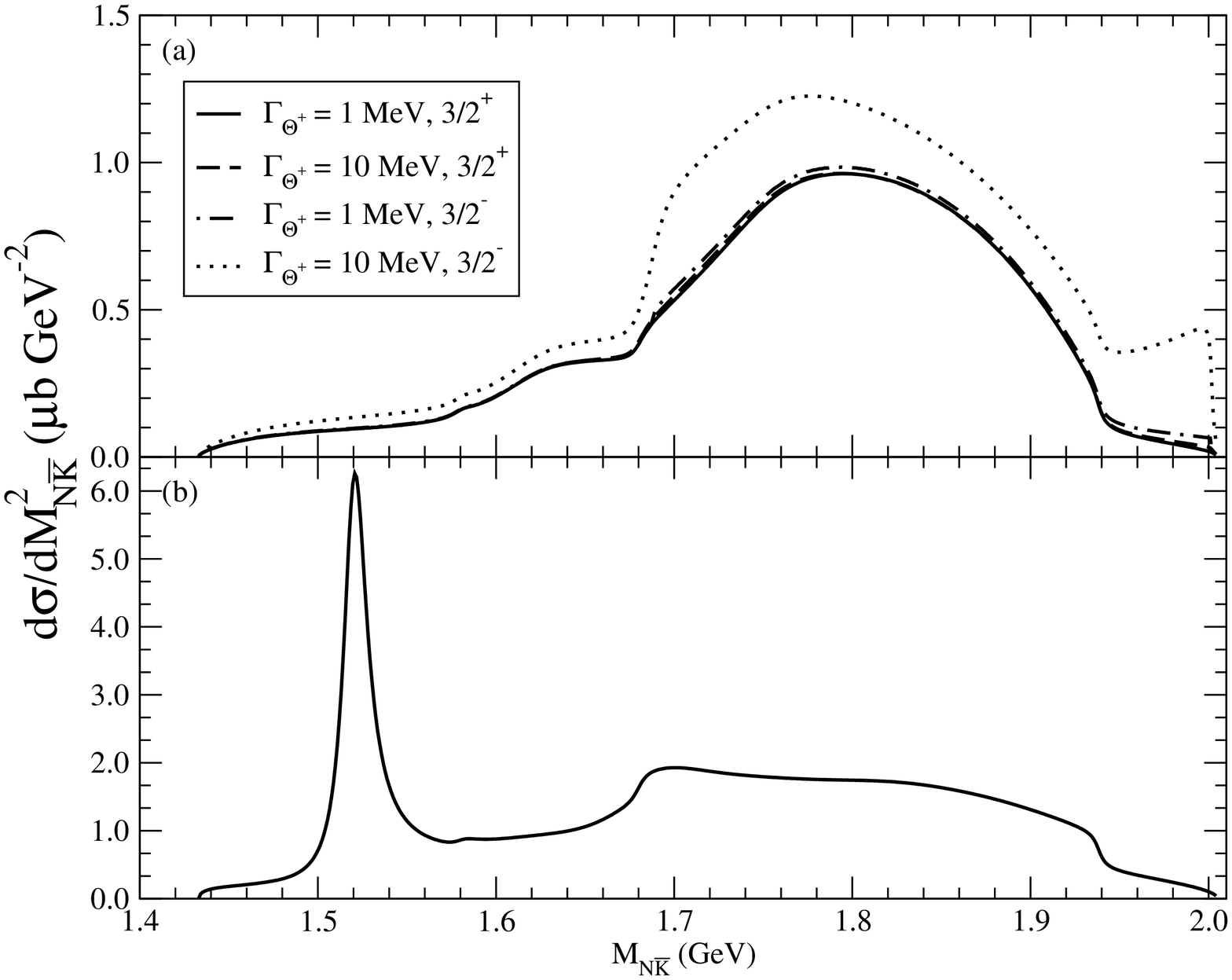}\hspace*{-0.2in}
\includegraphics[width=3in,angle=-90]{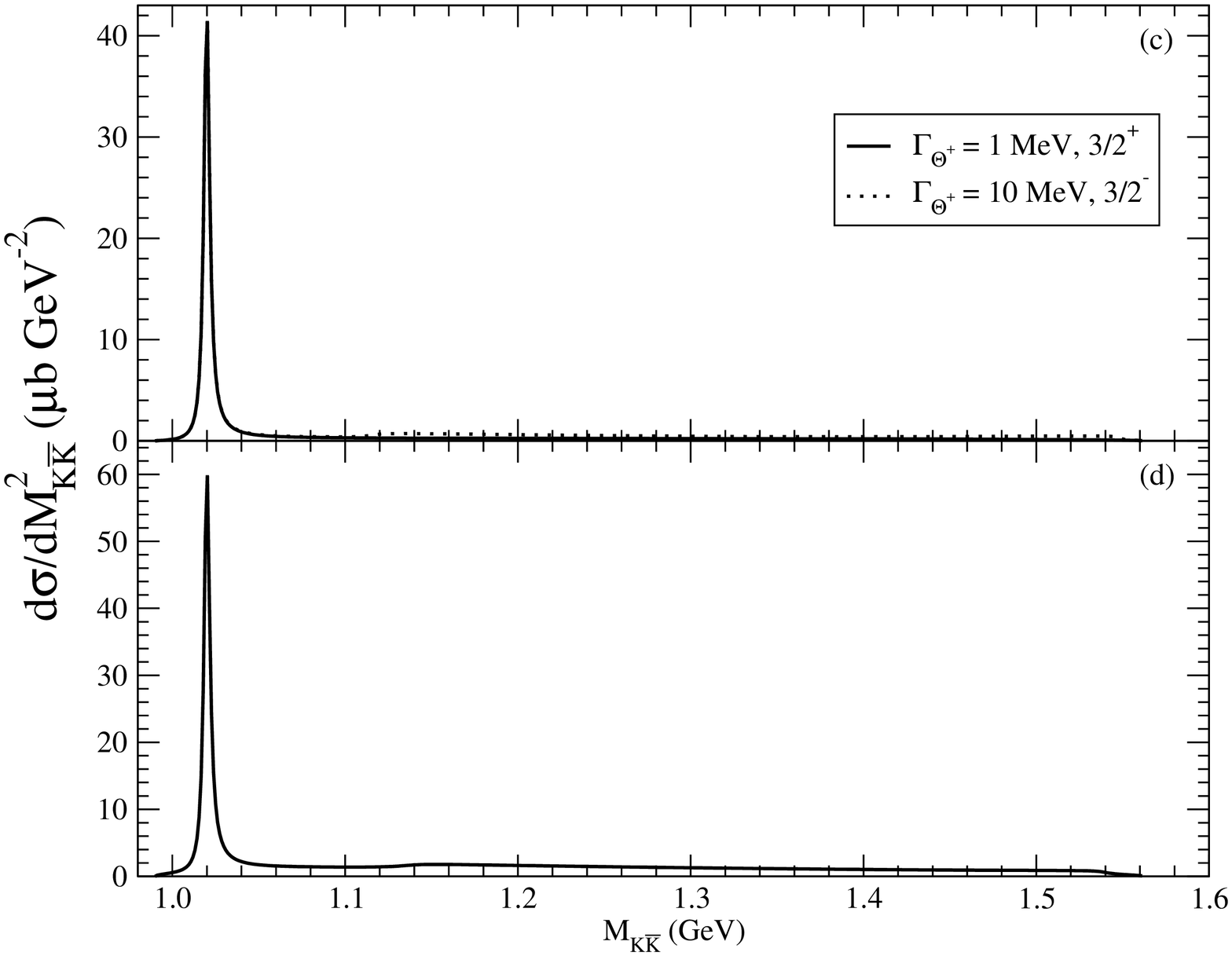}}
\caption{The differential cross section as functions of $M_{N\overline{K}}$
((a) and (b)), or $M_{K\overline{K}}$ ((c) and (d)). The curves in (a) and (c)
are for the process $\gamma n\to nK^+K^-$, while those in (b) and (d) are for
$\gamma p\to pK^+K^-$.\label{set2bc}} \end{figure}

Figure \ref{set2bc} shows the same differential cross section, but as functions
of different invariant masses. The curves in (a) and (b) show the differential
cross sections as a function of the mass of the nucleon-antikaon pair, while
the curves in (c) and (d) show it as a function of the mass of the
$K\overline{K}$ pair. In each case, the upper graph results from a neutron
target, while the lower graph is from a proton. Unlike the case with spin 1/2,
the contribution to the cross section of the $\Theta^+$ is now significant,
especially for the negative parity state, and gives rise to the strong
kinematic reflections seen in (a), and to a lesser extent, in (c) (the dotted
curves, for example).

\begin{figure}
\centerline{\includegraphics[width=3in,angle=-90]{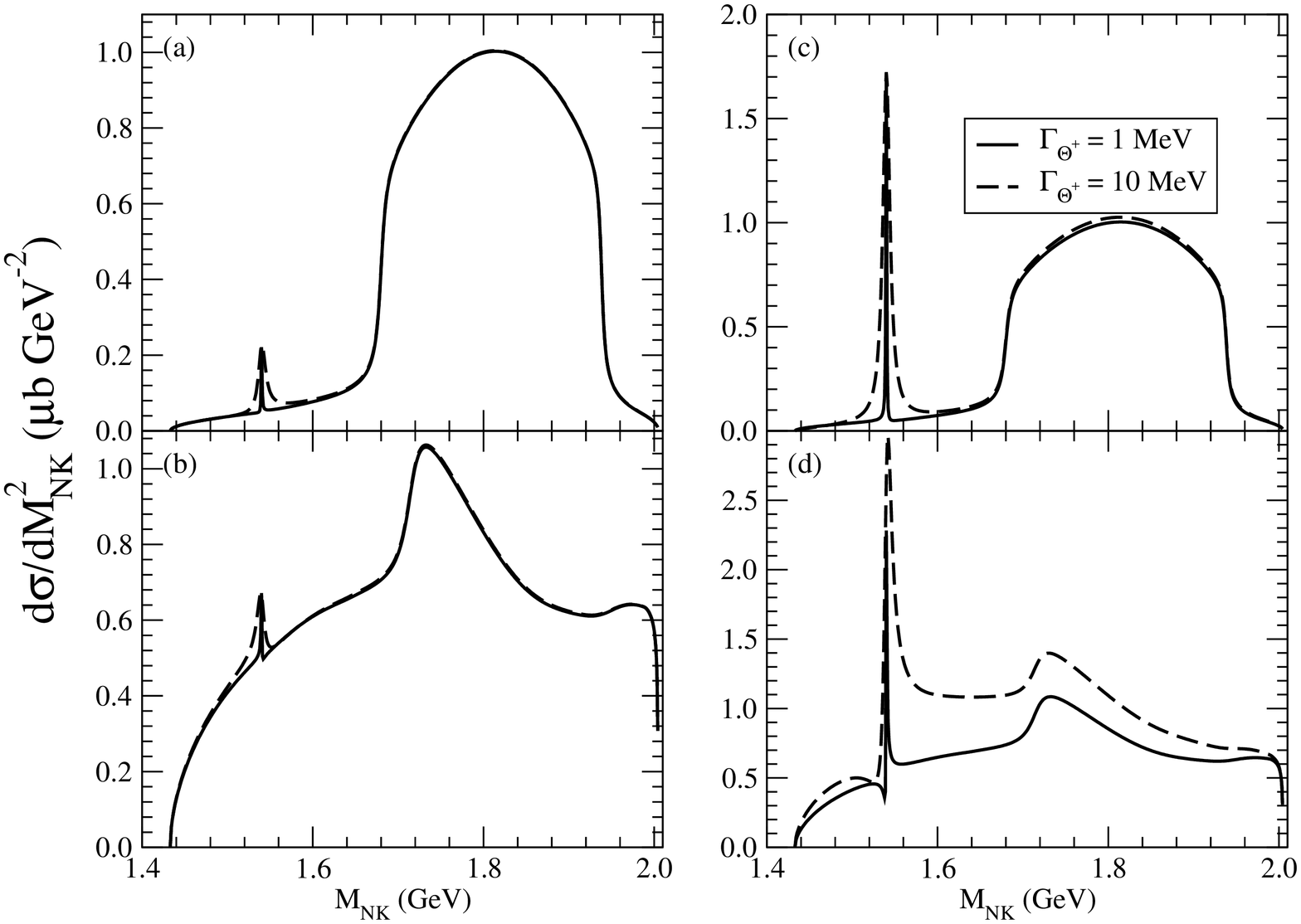}}
\caption{The differential cross section $\partial\sigma/\partial M_{NK}^2$
as a function of $M_{NK}$. The upper graphs are for the process $\gamma p\to
pK^0\overline{K}^0$, while the lower ones  are for $\gamma p\to
nK^+\overline{K}^0$.\label{set2d}} \end{figure}

Figure \ref{set2d} shows the differential cross section for the processes
$\gamma p\to p K^0\overline{K}^0$ ((a) and (c)) and $\gamma p\to n
K^+\overline{K}^0$  ((b) and (d)). The graphs on the left assume that the
$\Theta^+$ has positive parity, while those on the right have negative parity.
From the curves in (c) and (d), it should be clear that detecting a signal for
a $\Theta^+$ with $J^P=3/2^-$ would be relatively easy.

\subsubsection{Isovector, Spin-1/2 $\Theta^+$}

\begin{figure}
\centerline{\includegraphics[width=3in,angle=-90]{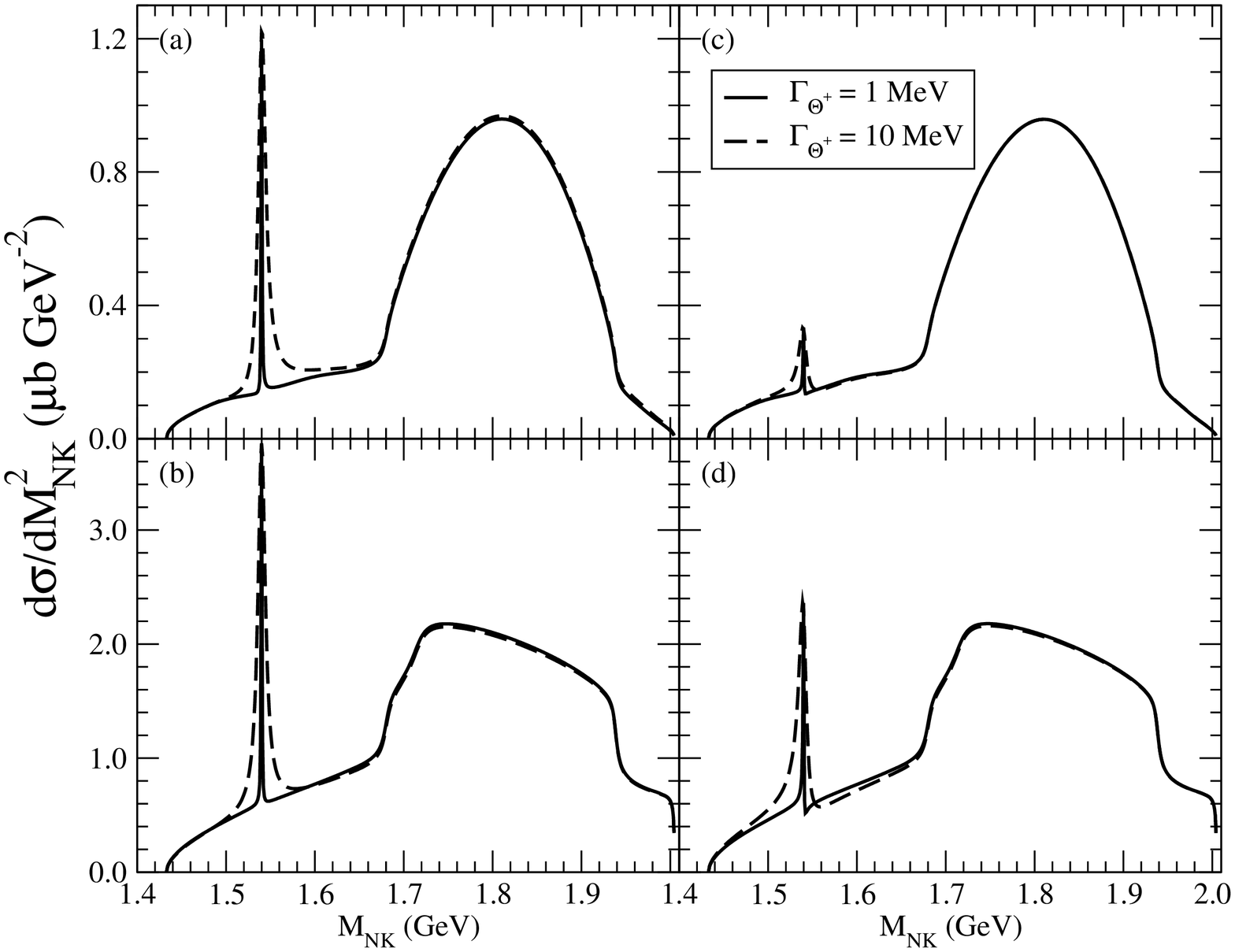}
\hspace*{-0.3in}\includegraphics[width=3in,angle=-90]{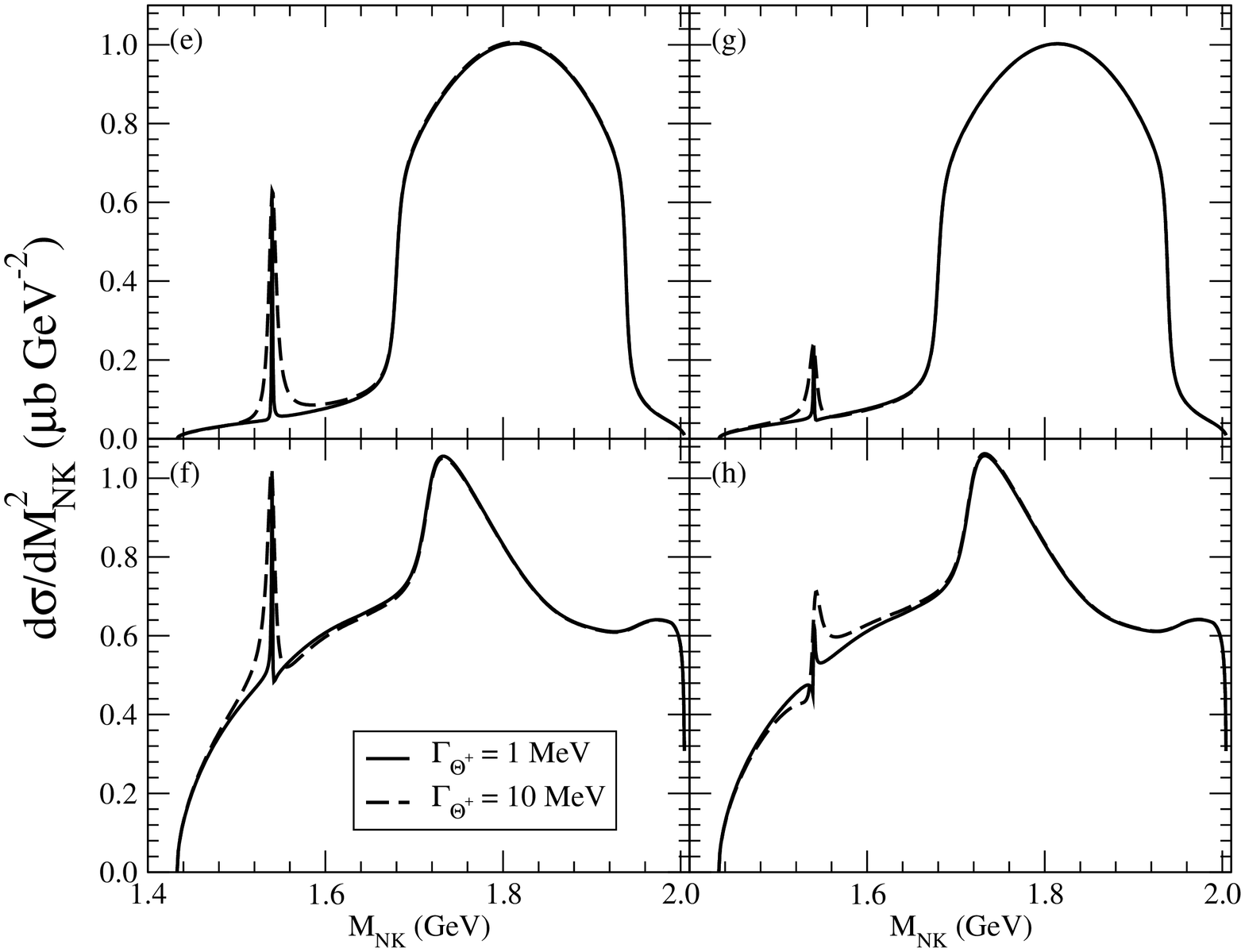}}
\caption{The differential cross sections $\partial\sigma/\partial
M_{NK}^2$ as a function of $M_{NK}$, for a spin-3/2 $\Theta^+$, for the processes $\gamma n\to
nK^+K^-$ ((a) and (c)), $\gamma p\to pK^+K^-$ ((b) and (d)), $\gamma p\to pK^0\overline{K}^0$ ((e) and
(g)) and $\gamma p\to nK^+\overline{K}^0$ ((f) and (h)). The curves in (a), (b), (e) and (f) all arise
from a positive-parity $\Theta^+$, while those in (c), (d), (g) and (h) all correspond to a $\Theta^+$
of negative parity.\label{set3ad}} 
\end{figure}

If the $\Theta^+$ were isovector, there would be a $\Theta^{++}$ state that
could be seen in $K^+ p$ final states, as well as a $\Theta^0$ that could be
present in $nK^0$ final states. Figures \ref{set3ad} (a) and (c) show the
effect of such a state in $\gamma n\to nK^+K^-$, while figures \ref{set3ad} (b)
and (d) show the effect in $\gamma p\to pK^+K^-$. The curves in (a) and (b)
assume that the $\Theta^+$ has positive parity, while those in (c) and (d)
assume that it has negative parity. In all cases, the spin is assumed to be
1/2. Figures \ref{set3ad} (e) and (g) show the differential cross section for
$\gamma p \to pK^0\overline{K}^0$, while (f) and (h) correspond to $\gamma p
\to nK^+\overline{K}^0$. (e) and (f) are for a $\Theta^+$ of positive parity,
while (g) and (h) assume that it has negative parity. The curves in (b) and (d)
suggest that a signal for a $\Theta^{++}$ should be comparable to that for a
$\Theta^+$, whatever the parity of the state. 

\subsection{Omitting $\Lambda(1520)$, $\phi(1020)$}

It is clear from the graphs shown in the preceding discussion that the
$\Lambda(1520)$ and the $\phi(1020)$ dominate the cross section for $\gamma N \to N
K\overline{K}$ for most channels. To enhance the possibility of isolating a
$\Theta^+$ signal, experimentalists impose kinematic cuts to eliminate the bulk of
the contribution from these two states. In our case, we will simple eliminate all
diagrams containing their contributions from the calculation. The curves that
result are presented in the next two subsections.

We note that we could also have imposed the same kinematic cuts on the model.
The (background) distributions that result when we do this are somewhat different from those
that we show, but the salient points of the discussion are unchanged.

\subsubsection{Isoscalar, Spin-1/2 $\Theta^+$}

\begin{figure}
\centerline{\includegraphics[width=3in,angle=-90]{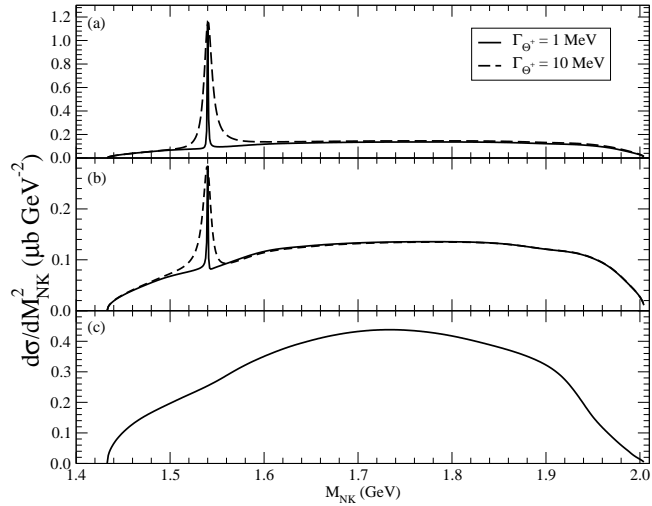}}
\caption{The differential cross section $\partial\sigma/\partial M_{NK}^2$
as a function of $M_{NK}$, for a spin-1/2 $\Theta^+$. The curves in (a) and (b) are
for the process $\gamma n\to nK^+K^-$, while (c) is for $\gamma p\to
pK^+K^-$. In addition, (a) is for a $\Theta^+$ of positive parity, while (b) is
for one of negative parity. \label{set4a}} \end{figure}

In figure \ref{set4a} we show the differential cross section that results for a
spin-1/2 $\Theta^+$, when the contributions of the $\Lambda(1520)$ and
$\phi(1020)$ are omitted from the calculation. In the case of a positive-parity
$\Theta^+$, a signal that may be easily identifiable results. In this figure,
the smooth background is provided by the non-exotic hyperons included in the
calculation. The graphs in (a) and (b) are for $\gamma n\to K^+K^-n$, while (c)
is for $\gamma p\to K^+K^-p$. The curves in (a) are for a $\Theta^+$ with
$J^P=1/2^+$, while those in (b) arise from a $\Theta^+$ with $J^P=1/2^-$.

\begin{figure}
\centerline{\includegraphics[width=3in,angle=-90]{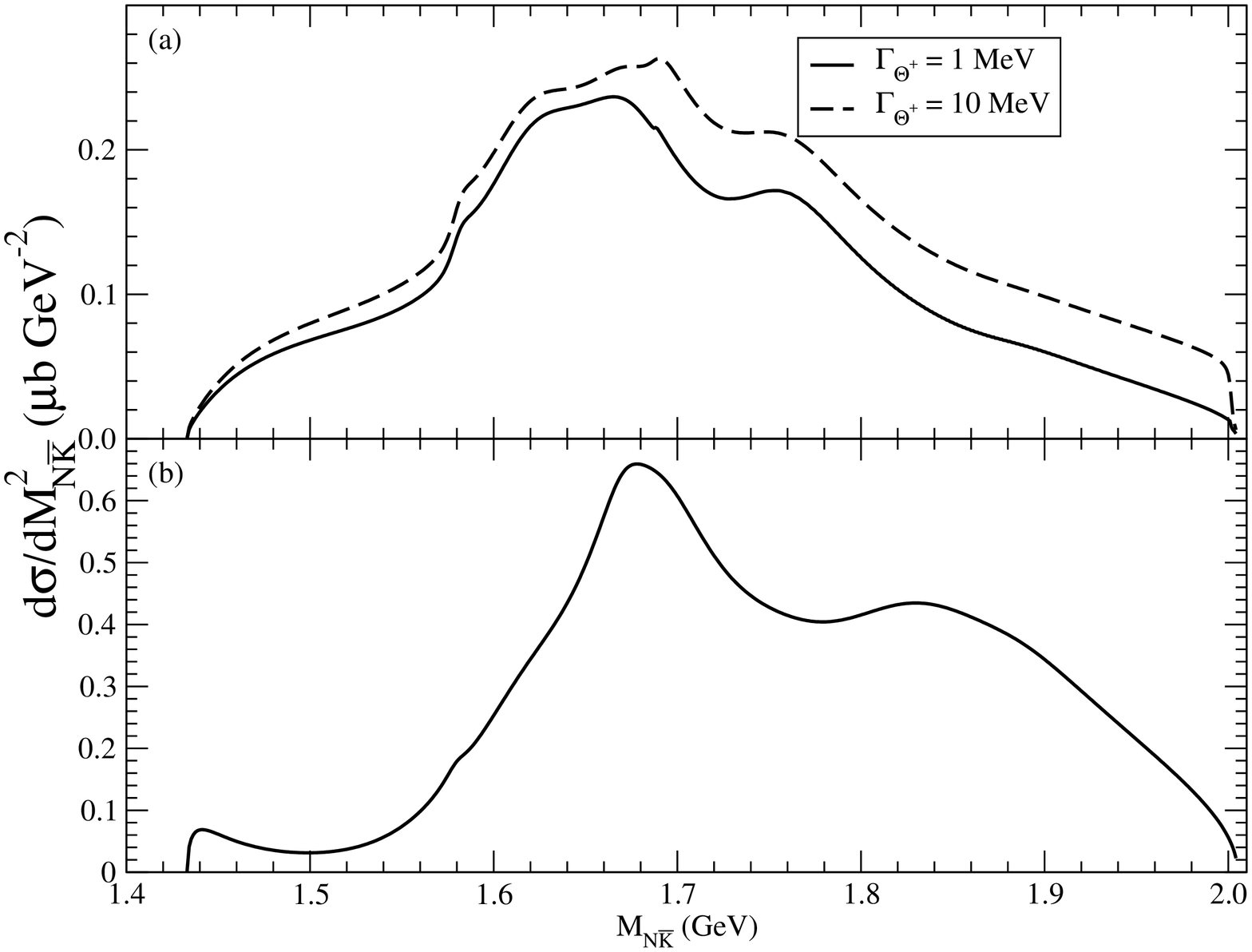}\hspace*{-0.2in}
\includegraphics[width=3in,angle=-90]{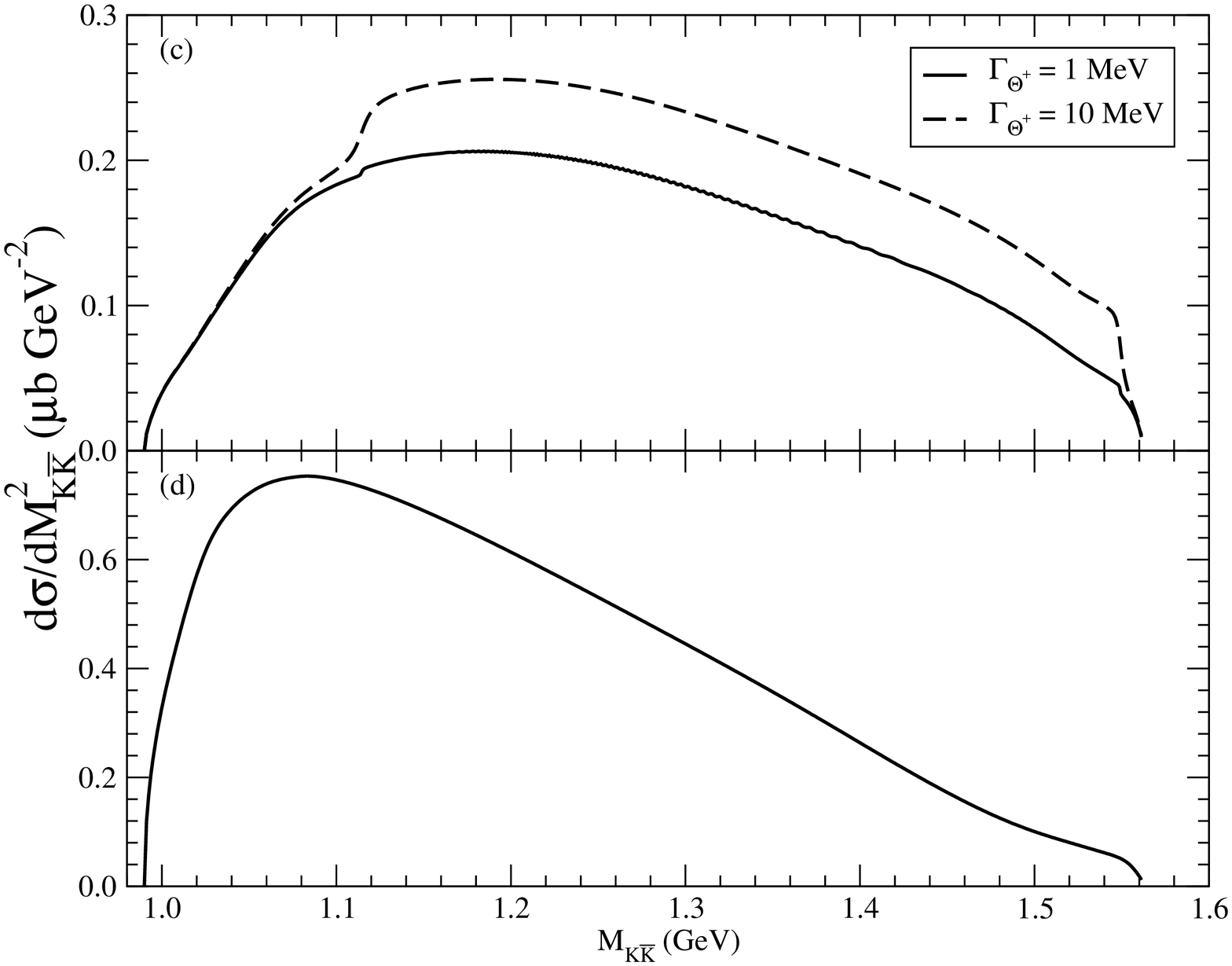}}
\caption{The differential cross section $\partial\sigma/
\partial M_{N\overline{K}}^2$ as a function of $M_{N\overline{K}}$ ((a) and 
(b)), and $\partial\sigma/\partial M_{K\overline{K}}^2$ as a function of 
$M_{K\overline{K}}$ ((c) and (d)). The curves in (a) and (c) are for the 
process $\gamma n\to nK^+K^-$, while those in (b) and (d) are for 
$\gamma p\to pK^+K^-$.\label{set4bc}}
\end{figure}

Figures \ref{set4bc} (a) and (b) show the differential cross section as a
function of the invariant mass of the $N\overline{K}$ pair. The effects of the
non-exotic hyperon resonances included in the calculation can be seen in these
curves. Figures \ref{set4bc} (c) and (d) show the same differential cross
sections as functions of the invariant mass of the $K\overline{K}$ pair. Since
there are no resonances left in this channel (in this model), relatively smooth
distributions with no prominent features result. In this figure, (a) and (c)
are for $\gamma n \to n K^+ K^-$, while (b) and (d) are for $\gamma p \to p K^+
K^-$. The shoulder seen near threshold in (b) results from the sub-threshold
$\Lambda(1405)$. If a larger coupling constant were chosen for this state, this structure would
be enhanced, while choosing a sufficiently smaller value will make this feature
disappear.

\begin{figure}
\centerline{\includegraphics[width=3in,angle=-90]{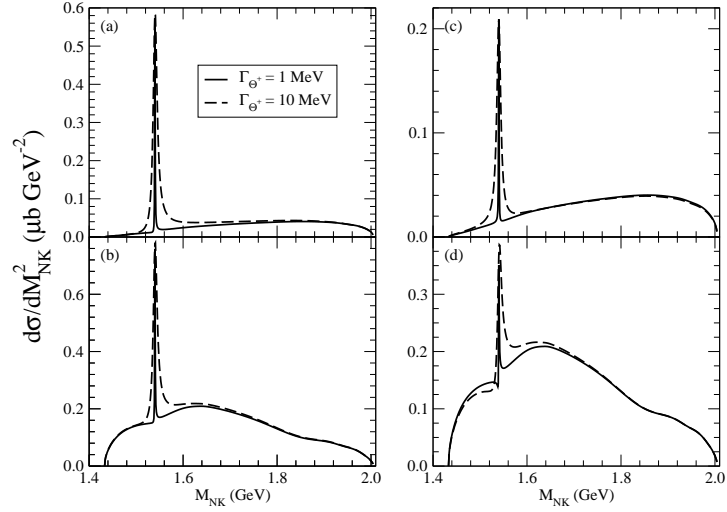}}
\caption{The differential cross section $\frac{\partial\sigma}{\partial
M_{NK}^2}$ as a function of $M_{NK}$. The curves in (a) and (c) are for the
process $\gamma p\to pK^0\overline{K}^0$, while those in (b) and (d) are for
$\gamma p\to nK^+\overline{K}^0$. (a) and (b) are for a positive parity
$\Theta^+$ while (c) and (d) are for a $\Theta^+$ with negative
parity.\label{set4d}}
\end{figure}

In figure \ref{set4d} we show the cross sections for the processes $\gamma p
\to p K^0 \overline{K}^0$ ((a) and (c)) and $\gamma p \to nK^+ \overline{K}^0$
((b) and (d)), in the channel that would show the isoscalar $\Theta^+$
resonance. The effects of the state are clearly seen, and suggest that for a
pentaquark of positive parity, either channel should provide a clear signal,
while for one of negative parity, the channel with two neutral kaons is better.

\subsubsection{Isoscalar, Spin-3/2 $\Theta^+$}

\begin{figure}
\centerline{\includegraphics[width=3in,angle=-90]{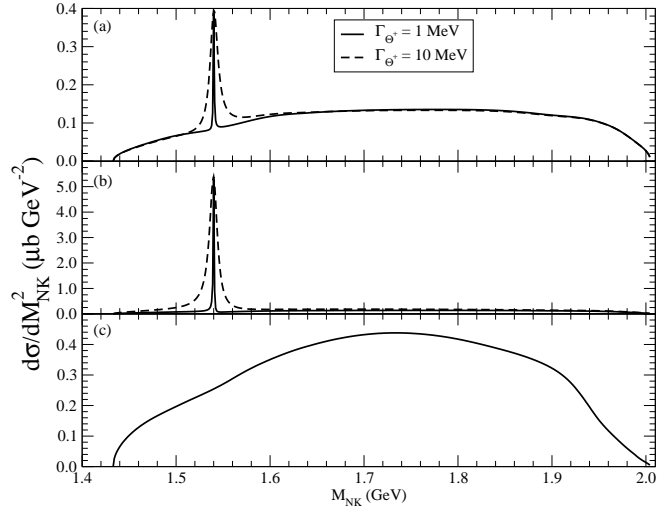}}
\caption{The differential cross section $\partial\sigma/\partial M_{NK}^2$ 
as a function of $M_{NK}$, for a spin-3/2 $\Theta^+$. The curves in
(a) and (b) are for the process $\gamma n\to nK^+K^-$, while (c) is
for $\gamma p\to pK^+K^-$. In (a), the $\Theta^+$ has positive parity, while in
(b), its parity is negative.\label{set5a}}
\end{figure}

Figure \ref{set5a} shows the differential cross section for a spin-3/2
$\Theta^+$ for the processes $\gamma n \to nK^+K^-$ ((a) and (b)), and $\gamma
p\to pK^+K^-$ ((c)). The curves in (a) assume that the $\Theta^+$ has positive
parity, while those in (b) assume negative parity. In the case of the negative
parity state, its contribution completely dominates the cross section. As
mentioned before, the signal generated by such a state should be comparable to
that generated by the $\Lambda(1520)$. The positive parity $\Theta^+$ also
provides a large signal above the `background', although it is not as dominant
as in the case of negative parity.

\begin{figure}
\centerline{\includegraphics[width=3in,angle=-90]{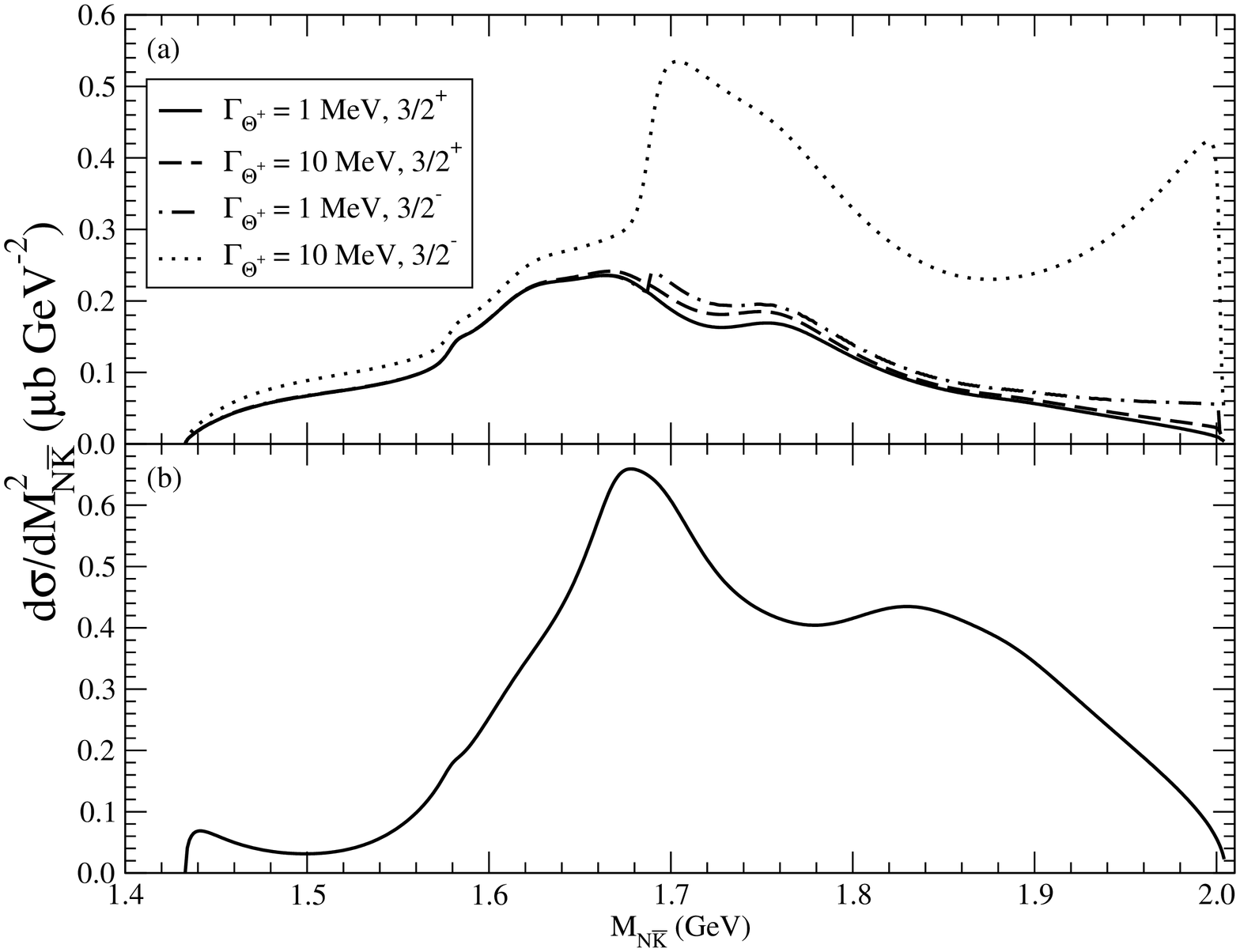}\hspace*{-0.2in}
\includegraphics[width=3in,angle=-90]{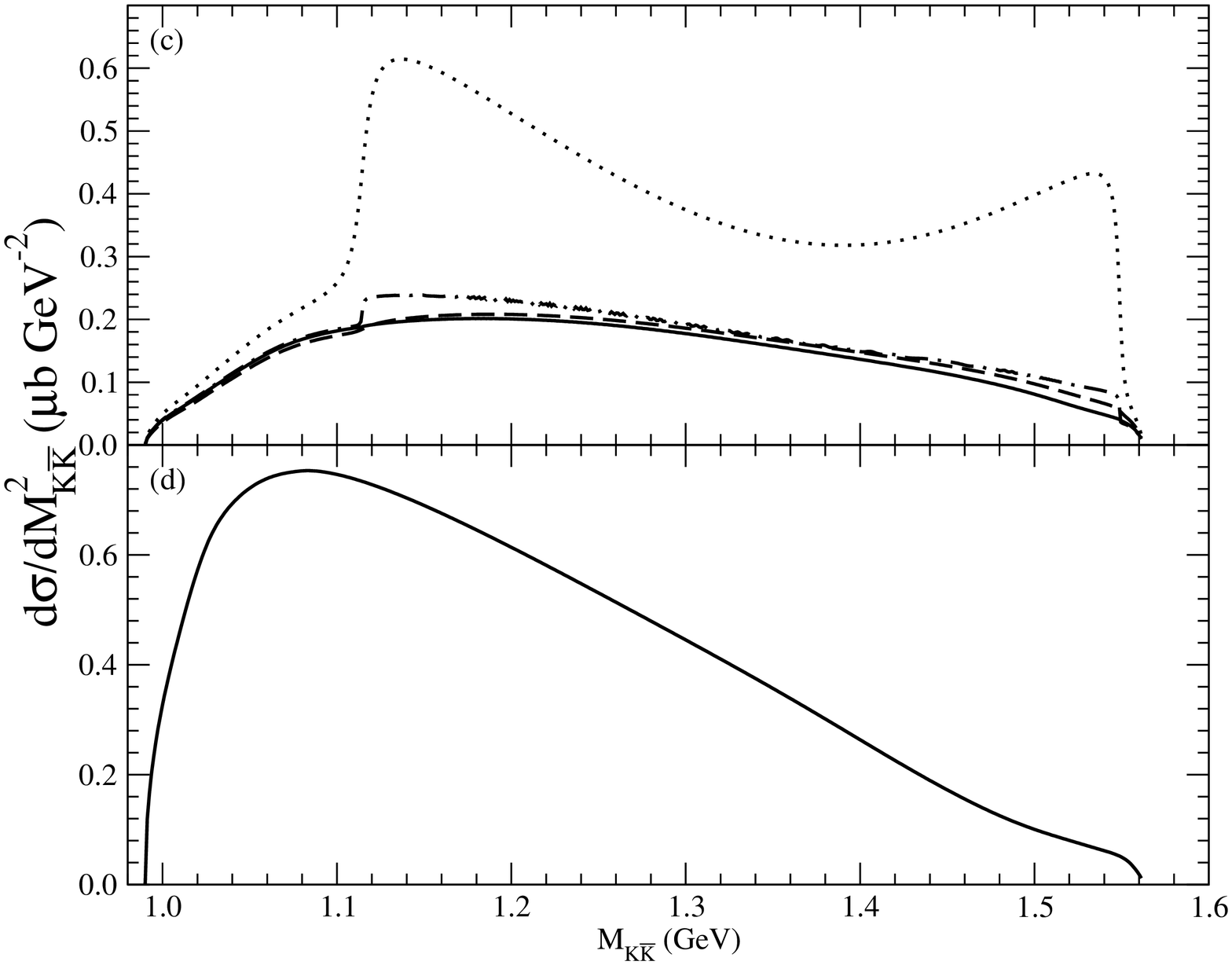}}
\caption{The differential cross section as functions of $M_{N\overline{K}}$
((a) and (b)), or $M_{K\overline{K}}$ ((c) and (d)). The curves in (a) and (c)
are for the process $\gamma n\to nK^+K^-$, while those in (b) and (d) are for
$\gamma p\to pK^+K^-$.\label{set5bc}}
\end{figure}

Figure \ref{set5bc} shows the same cross sections in terms of different invariant
masses. In (a) and (c) the large signal from the $\Theta^+$, particularly from the
negative parity version, show up as large kinematic reflections.

\begin{figure}
\centerline{\includegraphics[width=3in,angle=-90]{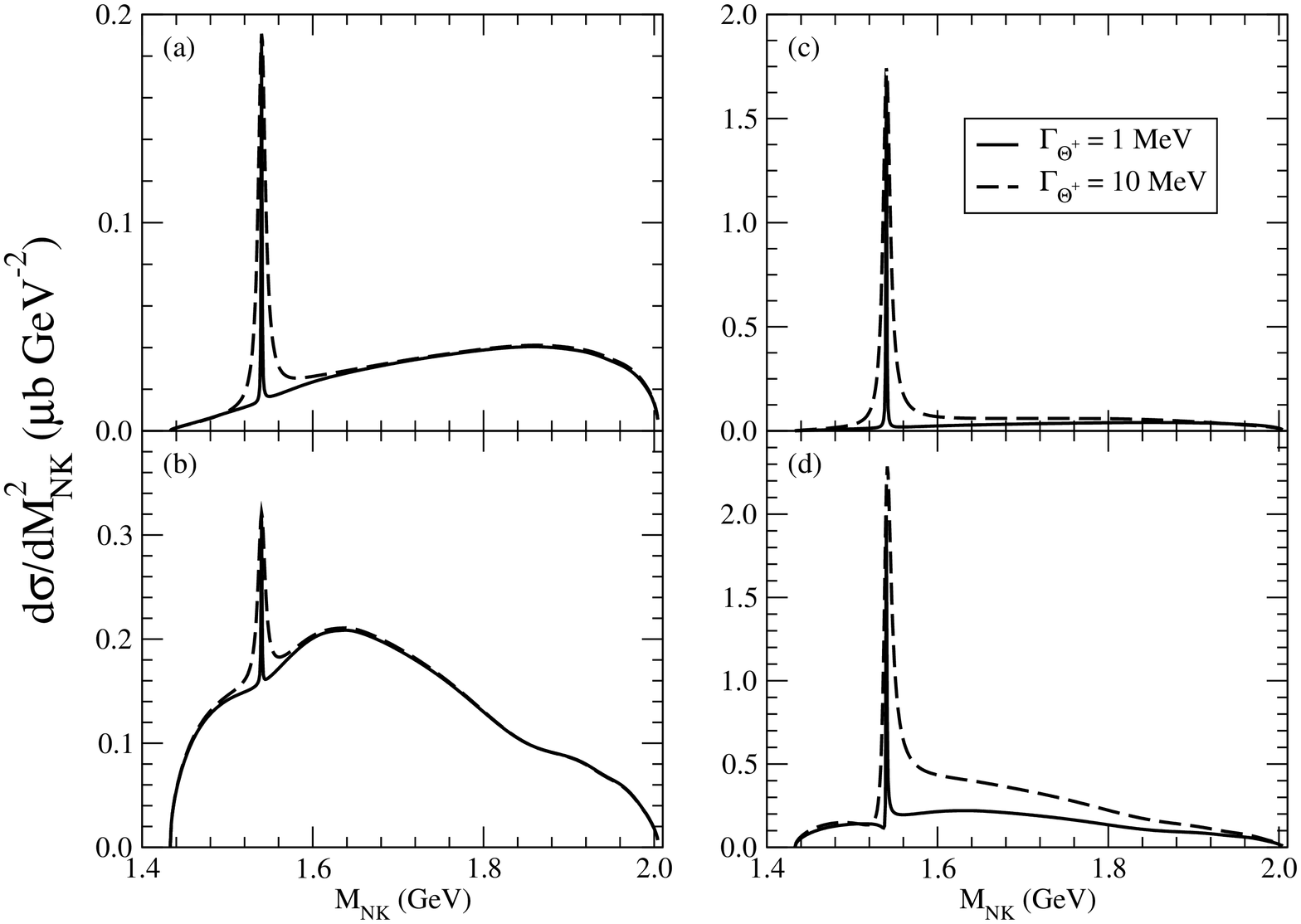}}
\caption{The differential cross section $\partial\sigma/\partial M_{NK}^2$
as a function of $M_{NK}$. The upper graphs are for the process $\gamma p\to
pK^0\overline{K}^0$, while the lower ones are for $\gamma p\to
nK^+\overline{K}^0$. Those on the left are for a $\Theta^+$ with positive parity, while those on the
right assume negative parity for the pentaquark. \label{set5d}}
\end{figure}

Figure \ref{set5d} shows the cross sections for $pK^0\overline{K}^0$ ((a)
and (c)) and $nK^+\overline{K}^0$ ((b) and (d)), both assuming a proton target.
In all cases, both for positive ((a) and (b)) and negative ((c) and (d))
parity, the model indicates that clear, easy-to-isolate signals should be
obtainable.

\subsubsection{Integrated Cross Sections}

In ref \cite{clas}, in $\gamma d\to npK^+K^-$, 212 events are in the peak for
the $\Lambda(1520)$, and there are 43 events in the peak of the $\Theta^+$. In
addition, the Saphir Collaboration \cite{saphir} report an estimated cross
section of 200 $nb$ for production of the $\Theta^+$ in the $\gamma p\to
nK^+\overline{K}^0$ channel. In this calculation, if we perform a numerical
integration around the peak of the $\Lambda(1520)$, we find that the cross
section in the peak in the channel $\gamma p\to pK^+K^-$ is of the order of 300
$nb$, with some small fraction of this arising from non-resonant contributions.
Table \ref{xsn} shows the integrated cross section under the peak of the
$\Theta^+$ for different channels in the different scenarios that we have
explored. In each case, the integration is performed from $M-2\Gamma$ to
$M+2\Gamma$, where $M$ is the mass of the $\Theta^+$ and $\Gamma$ is its
width. 

In each case there is some contribution arising from
`continuum' events that lie in the `right' kinematic regime. These continuum 
contributions represent a larger portion of the reported cross section for
$1/2^-$ pentaquarks than for $1/2^+$ pentaquarks, for instance. For a $1/2^-$
pentaquark with a width of 10 MeV, approximately half of the reported 18 $nb$
(in $\gamma n \to nK^+K^-$) arises from such continuum contributions. For a
pentaquark with the same quantum numbers but a width of 1 MeV, approximately
two-thirds of the reported 1.7 $nb$ in the same channel are from the continuum.

Assuming that the number of events seen is directly proportional to the cross
section, modulo questions of detector acceptances, resolution and efficiencies,
the JLab numbers suggest that the cross section for the $\Theta^+$ should be of
the order of 60 $nb$ around its peak in the channel $\gamma n\to nK^+K^-$. 

\begin{table}\caption{Total cross sections for production of the $\Theta^+$, in
different scenarios, for the channels in which it can be produced. The numbers in the
table are obtained from the versions of the model in which the $\phi$ and
$\Lambda(1520)$ are omitted. The second to fifth columns in the table
 correspond to a $\Theta^+$ with a width of 1 MeV, while the sixth to ninth 
 columns correspond to a width of 10 MeV. \label{xsn}} 
\begin{tabular}{|c|rrrr|rrrr|} 
Process & \multicolumn{4}{c|}{$\sigma\,\, (nb),\,\,\Gamma_{\Theta^+} = 1$
 MeV} &
\multicolumn{4}{c|}{$\sigma\,\, (nb),\,\,\Gamma_{\Theta^+} = 10$
 MeV} \\[+5pt] 
& $1/2^+$ &  $1/2^-$ &  $3/2^+$ &   $3/2^-$ &
  $1/2^+$ &  $1/2^-$ &  $3/2^+$ &   $3/2^-$ \\ \hline
$\gamma p\to pK^0\overline{K}^0$ & 2.6 & 1.0 & 0.9 & 7.7 & 25.3 & 9.6 & 8.9 &
73.6 \\ \hline 
$\gamma p\to nK^+\overline{K}^0$ & 4.3 & 2.4 & 2.3 & 10.0 & 44.8 & 27.9 & 26.0 &
110.5 \\ \hline 
$\gamma n\to nK^+K^-$ & 5.6 & 1.7 & 2.2 & 24.0 & 54.5 & 18.0 & 22.8 & 229.9 \\ \hline 
$\gamma n\to pK^0K^-$ & 5.9 & 1.8 & 2.0 & 24.0 & 56.1 & 19.0 & 21.2 & 225.1 \\ \hline
\end{tabular} 
\end{table}

The numbers in the table indicate that the versions of the model discussed so
far are inconsistent with the $\Theta^+$ signal measured at Jefferson Lab, for
instance. The estimated cross section inferred for production of the $\Theta^+$
in the process $\gamma n\to nK^+K^-$ is about 60 $nb$, and for a
pentaquark with $J^P=1/2^+$ and a width of 10 MeV, the cross section calculated
in this model is 55 $nb$. However, such a large width for the state appears to
be in contradiction to cross sections observed in other processes
\cite{nussinova} - \cite{gibbs}: consistency with such observations would
dictate that the preferred scenario is for a pentaquark with a width of 1 MeV.
In this case, the scenario that most closely matches JLab observations is that
with a $3/2^-$ pentaquark. However, the results of this calculation suggests
that such a state should not need kinematic cuts for observation. None of the
scenarios with the narrower $\Theta^+$ match the reported Saphir cross section
of 200 $nb$.

We have examined cross sections for these processes at smaller values of
$\sqrt{s}$. While the overall cross sections change, the relative strengths of
various contributions remain similar to what they are at $\sqrt{s}=2.5$ GeV. In
particular, the ratio of integrated cross section for the $\Theta^+$ and the
$\Lambda(1520)$, remains similar to what it is at $\sqrt{s}=2.5$ GeV. Thus, the
discrepancy between the results of our model and the signal seen at JLab would
remain as difficult at lower energies.

\subsection{The Role of the $K^*$}

The preceding subsections suggest that, apart from the case of a $\Theta^+$
with $J^P=1/2^-$, a signal for this pentaquark should be readily observable,
especially when the $\Lambda(1520)$ and $\phi$ are omitted from the
calculation. However, there is still an inconsistency between what we have
shown and what has been observed experimentally at JLab. 

In the results presented, the contributions of the $K^*$ mesons have been
limited to diagrams in which they couple only to ground state hyperons and
nucleons. At this point, there are no contributions in which the $K^*$ couple
to excited hyperons, nor to the $\Theta^+$. It is relatively easy to include
such contributions, and in so doing, we can increase the cross section for
production of the $\Theta^+$.

The phenomenological Lagrangian for the coupling of the $\Theta^+$ to the $K^*$ may be
written 
\begin{equation} 
{\cal L}= \overline{N}\left(G_v^{K^*N\Theta^+} \gamma^\mu K^*_\mu
+\frac{G_t^{K^*N\Theta^+}}{M_N+M_{\Theta^+}} \gamma^\mu \gamma^\nu \left(\partial_\nu
K^*_\mu\right)\right)\Theta +  H. c., 
\end{equation}  
if the $\Theta^+$ is assumed to have $J^P=1/2^+$. This is the only scenario
we discuss. The two coupling constants $G_v^{K^*N\Theta^+}$ and
$G_t^{K^*N\Theta^+}$ are unknown. In table \ref{xsnks} we show results for
different values of the vector coupling constant, with the tensor coupling set
to zero. We see that relatively modest values of the vector coupling are 
sufficient to
give a cross section of about 60 $nb$ for production of the $\Theta^+$ in the
$\gamma n\to nK^+K^-$ channel. However, even that modest value for the coupling
constant (of about 6) is
somewhat larger than values postulated by some authors. For instance, Close and
Zhao \cite{close} have suggested that $\left(G_v^{K^*N\Theta^+}\right)^2\approx 3$. In other models, similar
values have been used. With $G_v^{K^*N\Theta^+}=6$, our model predicts a very
large cross of 280 $nb$ for the production of the $\Theta^+$ in the $\gamma p\to
pK^0\overline{K}^0$ channel, and a cross section of 73 $nb$ in $\gamma p\to
nK^+\overline{K}^0$. This mean means that, in this scenario, the contribution of the $K^*$ dominates
the production of the $\Theta^+$. These numbers indicate that the reported JLab
results are not consistent with the estimated Saphir cross section of 200 $nb$. In addition, in $pK^0\overline{K}^0$, the
predicted cross section for production of the $\Theta^+$ is comparable to that
for production of the $\Lambda(1520)$, implying that the signal should be easily
observable.

\begin{table}[h]\caption{Total cross sections for production of the $\Theta^+$, in
different scenarios, for different processes. The numbers in the
table are obtained from the versions of the model in which the $\phi$ and
$\Lambda(1520)$ are omitted. All numbers assume that the $\Theta^+$ has
$J^P=1/2^+$ and a width of 1 MeV. The tensor coupling is set to zero, and four
different values of the vector coupling are used. All channels in which the
$\Theta^+$ can be produced as a resonant state are shown. \label{xsnks}} 
\begin{tabular}{|l|r|r|r|r|} 
Process & $\sigma$ $(nb)$, $G_v^{K^*N\Theta^+}=2$ & $\sigma$ $(nb)$,
$G_v^{K^*N\Theta^+}=4$ &
$\sigma$ $(nb)$, $G_v^{K^*N\Theta^+}=6$  & $\sigma$ $(nb)$,
$G_v^{K^*N\Theta^+}=8$ \\ \hline 
$\gamma p\to p K^0\overline{K}^0$ & 32.7 & 125.9 & 282.2 & 502.8 \\ \hline 
$\gamma p\to n K^+\overline{K}^0$ & 11.5 & 34.6 & 73.4 & 128.1 \\ \hline 
$\gamma n\to n K^+K^-$ & 10.0 & 30.2 & 66.1 & 117.8 \\ \hline 
$\gamma n\to p K^0 K^-$ & 30.7 & 118.6 & 269.7 & 484.0 \\ \hline
\end{tabular} \end{table}

\section{Summary and Outlook}

We have examined the process $\gamma N\to K\overline{K}N$ within the framework
of a phenomenological Lagrangian. We have examined a number of scenarios for
pentaquark production, and have found that the largest production cross section
occurs for a $\Theta^+$ with $J^P=3/2^-$. However, in such a scenario, the cross
section for its production is comparable to that for production of the
$\Lambda(1520)$, and kinematic cuts should probably not be needed to enhance the
signal. 

If the $\Theta^+$ has $J^P=1/2^+$, the cross section for its production is
significantly less than that for production of the $\Lambda(1520)$, by almost
two orders of magnitude if its width is of the order of 1 MeV. This means that special mechanisms are required to
account for the number of events seen in the JLab experiment, relative to the
$\Lambda(1520)$. One possibility is that the coupling of the pentaquark to the
$K^*$ is large, so that the dominant mechanism of production involves the
$K^*$. However, this then leads to a signal for which kinematic cuts should not
be necessary. One can also invoke the couplings of a number of $N^*$ resonances,
but the conclusion about the size of the signal would remain unchanged.

The only scenario (that we can think of) that would give the appropriate ratio
between the cross section for production of the $\Lambda(1520)$ and the
$\Theta^+$ is for the production of the $\Lambda(1520)$ to be suppressed
even further than the suppression we have already obtained through the use of
form factors. However, this seems unlikely, as the calculated cross section for
producing this state is of the same order of magnitude as those published by
Barber {\it et al.} \cite{barber}.

\subsection{Outlook}

This calculation is not without its shortcomings. The most important
shortcoming is the fact that a very simple prescription has been employed to
regulate the high-energy behavior of the model. A more realistic treatment,
consistent with the requirements of gauge invariance, will have to be
implemented before such a calculation is applied to other processes in the
future. 

There are prospects for measuring a number of final states with two
pseudoscalar mesons at JLab and at other facilities. In particular, there are
on-going analyses of the processes $\gamma N\to\pi\pi N$, $\gamma N\to KK\Xi$,
$\gamma N\to K\pi\Lambda$ and $\gamma N\to K\pi\Sigma$. The calculation we have
presented has been set up in such a way that it may be applied to any of these
(or other) processes in a relatively straightforward manner. The core of the
code was originally generated for $\gamma N\to\pi\pi N$, and the modifications
necessary for $\gamma N\to K\overline{K}N$ were not overly difficult. Thus, we
may expect to apply the methods used herein to other processes in the
not-too-distant future.

\section*{Acknowledgment}

The author thanks J. L. Goity, J. - M. Laget and F. Gross for reading the
manuscript, and for discussions. This work was supported by the Department of
Energy through contract DE-AC05-84ER40150, under which the Southeastern
Universities Research Association (SURA) operates the Thomas Jefferson National
Accelerator Facility (TJNAF).

\end{document}